\documentclass[%
 reprint,
 amsmath,amssymb,
 aps,
]{revtex4-1}

\usepackage{graphicx}
\usepackage{dcolumn}
\usepackage{bm}

\usepackage{graphics}
\usepackage{bm}
\usepackage{amsmath}
\usepackage{epstopdf}
\usepackage{amsmath}
\usepackage{amsfonts}
\usepackage{color}
\usepackage[]{algorithm2e}
 \usepackage{tikz}
 \usepackage{mathtools}
 \usepackage[percent]{overpic}
 \usepackage{float}
 \usepackage{subfig}
 \usepackage{comment}
 \usepackage{cleveref}

\let\MYoriglatexcaption\caption
\renewcommand{\caption}[2][\relax]{\MYoriglatexcaption[#2]{#2}}


\renewcommand{\eqref}[1]{(\ref{#1})}

\begin{document}

\preprint{APS/123-QED}

\title{SEAIR epidemic spreading model of COVID-19} 

\author{Lasko Basnarkov$^{1,2}$}
\email{lasko.basnarkov@finki.ukim.mk}
\affiliation{$^{1}$Faculty of Computer Science and Engineering, SS. Cyril and Methodius University, P.O. Box 393, 1000 Skopje, Macedonia\\}%

\affiliation{$^{2}$Macedonian Academy of Sciences and Arts, P.O. Box 428, 1000 Skopje, Macedonia}%


\date{\today}

\begin{abstract}

We study Susceptible-Exposed-Asymptomatic-Infectious-Recovered (SEAIR) epidemic spreading model of COVID-19. It captures two important characteristics of the  infectiousness of COVID-19: delayed start and its appearance before onset of symptoms, or even with total absence of them. The model is theoretically analyzed in continuous-time compartmental version and discrete-time version on random regular graphs and complex networks. We show analytically that there are relationships between the epidemic thresholds and the equations for the susceptible populations at the endemic equilibrium in all three versions, which hold when the epidemic is weak. We provide theoretical arguments that eigenvector centrality of a node approximately determines its risk to become infected.









\end{abstract}

\maketitle

\section{Introduction}

Understanding epidemic spreading of contagious diseases and effectiveness of various countermeasures is of high interest for the public health and the society in general, with important contributions provided by epidemiologists, mathematicians and physicists as well. Although earliest theoretical work in the mathematical epidemiology dates back to Daniel Bernoulli \cite{bernoulli1760essai}, the development of the modern approach started in the beginning of the past century \cite{hamer1906epidemic, ross1911prevention,  kermack1927contribution}. 
In the last two decades, since the emergence of the complex networks theory, epidemic modeling has gained novel insights. By modeling the contacts between the individuals with complex networks, some associations were found between the epidemic threshold and the network properties like the degree distribution or the leading eigenvalue of the adjacency matrix \cite{pastor2001PRE, pastor2001PRL, newman2002spread,   vespignani2012modelling, pastor2015epidemic,  de2018fundamentals}. Moreover, the epidemic spreading has grown as a concept that extends its original design for modeling diffusion of infectious diseases to sharing ideas, rumors, or computer viruses \cite{pastor2015epidemic}. 

In the classical approach, the individuals are conveniently grouped in compartments or classes. The mathematical models for epidemic spreading in this setting are systems of differential equations for the evolution of the size of those compartments. In such models, there is an assumption of homogeneous mixing, which means that the pathogen can spread between each pair of individuals with equal probability. In a more realistic modeling each individual is considered as node in certain network of contacts, where infection can spread only among neighbors in that network. This is particularly relevant, because real world networks besides being random, among others they possess properties like small-world phenomenon \cite{watts1998collective}, or scale-free distribution of the node degrees \cite{barabasi1999emergence}. One approach for studying the disease spreading on networks is the heterogeneous mean-field \cite{pastor2001PRL, boguna2003absence} in which all nodes with the same degree are assumed to be statistically equivalent. The quenched mean-field technique is applied in even more realistic scenario, where each node is treated separately \cite{wang2003epidemic, van2008virus}. Among major contributions in the field of epidemic spreading on networks one should mention the studies of disease localization \cite{goltsev2012localization, de2017disease}, spreading on multilayer networks \cite{, de2017disease, wang2020epidemic}, the assumption of non-exponential distribution of periods between consecutive events \cite{van2013non, starnini2017equivalence} and the effects of the delay in the recovery \cite{xia2013effects} and in the infection \cite{xia2012sir}. 

Various infections are characterized with different stages in the course of development of the disease in the host, starting from contracting the pathogen to the healing. Depending on the disease under study, several compartments, or classes are defined in order to differentiate between the stages. The most frequently used compartments are the Susceptible (S), Exposed (E), Infectious (I) and Recovered (R) \cite{hethcote2000mathematics, pastor2015epidemic}. The meaning of the compartmental symbols usually is: S -- healthy individuals subjected to infection, E -- infected which do not transmit the disease yet, I -- infected and infectious and R -- cured which cannot become infected again. The most popular models are the SIS and SIR, which are sufficiently simple to provide mathematical tractability, and powerful enough to capture the features of epidemics of many contagious diseases \cite{hethcote2000mathematics}. 


In this theoretical work we consider the known Susceptible-Exposed-Infected-Recovered (SEIR) model augmented with another state, the asymptomatic state (A), which precedes the infectious. The chosen SEAIR model has a built-in delay which means that the infected person do not start to spread the disease immediately, allows to address the different contagiousness of the disease in different phases and captures the presence of undetected spreaders. Also, it is simple enough to allow for theoretical study of different properties. We study the SEAIR model with two approaches: the classical, which uses differential equations, and the one based on statistical physics framework, in which is considered discrete-time epidemic model on complex networks. We study the latter model on random regular graphs and on complex networks separately. For the three versions we obtain the epidemic threshold, equation for the fraction of susceptible individuals at the end of epidemic and we study the linear stability of the disease-free and endemic equilibria. The results for the random regular graphs hold when the contagiousness is weak, while for the complex networks it is also needed that epidemic is small-scale one in a sense that it can affect a tiny fraction of the population. We furthermore study the roles of the leading eigenvalue and the principal eigenvector of the adjacency matrix in the spreading. It is already known that the leading eigenvalue determines the epidemic threshold \cite{wang2003epidemic, prakash2010got}. We show that the principal eigenvector and its associated eigenvector centrality have important role in estimation of the risk of infection. We finally note that, the techniques which are applied for analysis of the Jacobian matrix at the equilibria might be relevant in studies of linear stability of coupled multidimensional dynamical systems.


The paper is organized as follows. First, in Section \ref{SEC:Covid-models} we present some of the relevant literature about epidemic spreading models of COVID-19 and their relation to the proposed model. Then, in Section \ref{SEC:compart} we introduce and analyze the SEAIR compartmental model. In the following Section \ref{SEC:Discrete} is studied the discrete-time version on random regular graphs and complex networks, while in Section \ref{SEC:Numerical} we present some results of the numerical experiments. The paper ends with the Conclusion.

\section{Epidemic spreading models of COVID-19 and their relation to the SEAIR model} \label{SEC:Covid-models}

The ongoing COVID-19 is the largest pandemic in modern history. The understanding of the virus and its influence on the infected individuals are still in progress. However, it was established that a key feature is the possibility that an infected person can spread the disease before onset of symptoms or without having them at all \cite{he2020temporal, du2020serial}. Another important observation is that the median incubation period of the disease is approximately five days \cite{lauer2020incubation}. Thus, the asymptomatic spreaders and individuals in the incubation period are hidden disease carriers. Their presence poses a challenge in the control of the epidemic, in the planning of healthcare capacity buildup or the relaxation of lockdown measures, and even for estimation of the population affected by the pathogen. The mathematical models of the epidemic spreading are adapted accordingly to capture the key features of the spreading of the COVID-19, for which the most basic and popular ones, SIS and SIR are not satisfactory. The existence of incubation period is addressed by including compartment of exposed individuals (E), which in certain studies in the literature is denoted as latent. The absence or presence of symptoms is addressed in different ways in the literature. In one approach \cite{di2020expected}, it is considered presence of presymptomatic, or prodromic phase, where the individual is already infectious, which is followed by either asymptomatic or symptomatic phase which can have three different severity levels. In another study \cite{aleta2020modeling},
there are separate latent compartments for the asymptomatic and symptomatic individuals, and it is considered different contagiousness by the infected individuals depending on the presence or absence of symptoms. The observation of existence of super spreaders of the disease was also included in a mathematical model \cite{ndairou2020mathematical}. In the analysis performed in \cite{giordano2020modelling}, the infected persons which are detected, either asymptomatic or symptomatic, are accounted in separate class from those that are not. Since the quarantine is one of the key defence measures against the spreading, the quarantined compartment was considered as well in many works \cite{zhao2020modeling, peng2020epidemic, gatto2020spread, nabi2020forecasting}. Also, to account for the disease outcomes and the burden on the healthcare capacities were introduced compartments for the hospitalized, for those under intensive care, and for those which did not survive the disease \cite{aleta2020modeling, gatto2020spread, di2020expected, nabi2020forecasting}.

From one side, the simpler models are easier to study and in general provide better estimation of the model parameters which are needed for making predictions for future development of an epidemic. From another side, the more complex models obviously allow for better description of certain features of particular epidemic like the COVID-19. Thus, one has to make a choice about what kind of model to apply in order to obtain particular insight. In this work, the aim is to work with a model that will address the key features of COVID-19 epidemic, the latency and the existence of asymptomatic transmission, and study it in frames of complex networks theory which accounts for nontrivial pattern of contacts between the individuals. Accordingly, we separate the disease transmitters in two compartments. In the asymptomatic one is included an individual which do not have symptoms at the moment of transmission of the virus, regardless whether he or she will obtain them later or not. In the infectious compartment are accounted the persons which can infect the others and have symptoms already. Thus, in the asymptomatic and infectious compartments are included the three kinds of disease spreaders: always asymptomatic, currently asymptomatic which will develop symptoms later and symptomatic. More detailed description of the SEAIR model are given in the following two sections. We finally note that the proposed model can be considered as similar to that in  \cite{liu2020covid}, where the individuals which are not detected as carriers of the virus are considered as unreported to the authorities and can transmit the disease, while the reported, or detected ones, do not contribute to the spreading. Another model closely related to SEAIR is that in \cite{arino2020simple}, because it also has incorporated pre-symptomatic transmission and two different compartments for the asymptomatic and symptomatic individuals.

\section{SEAIR compartmental model}\label{SEC:compart}
 
 Let the variables $S$, $E$, $A$, $I$ and $R$ denote the fraction of individuals which are respectively susceptible, exposed, asymptomatic, infectious and recovered. We assume that the exposed state corresponds to the incubation period when a host has the pathogen, but he or she cannot infect the others. In the asymptomatic state the individual spreads the disease, possibly with higher virulence without being aware about having the virus. The person can even recover without ever noticing that he, or she had the disease. Certain percentage of carriers of the virus will show symptoms and we classify them as infectious. Let the infecting rate of the asymptomatic persons be $\alpha$, while of infectious ones be $\beta$. The rate at which exposed individuals become asymptomatic is $\gamma$. The growth of the fraction of infectious hosts, which have symptoms is determined with rate $\sigma$. For simplicity, we assume that healing of both the asymptomatic and infectious persons is modeled with the same rate $\mu$. We emphasize that $\mu$ does not exactly correspond to the time of complete healing, but the period in which a person can infect the others. With these assumptions one has the following SEAIR compartmental model
 \begin{eqnarray}
    \Dot{S} &=&  - \alpha A S -\beta I S,\nonumber\\
    \Dot{E} &=& \alpha A S + \beta I S - \gamma E, \nonumber\\    
    \Dot{A} &=& \gamma E - \sigma A - \mu A, \nonumber\\    
    \Dot{I} &=& \sigma A - \mu I, \nonumber\\       
    \Dot{R} &=& \mu A + \mu I.
    \label{eq:SEAIR_comp_def}
\end{eqnarray}
We have neglected the births and deaths in the population and one can easily verify that the total number of persons in all states is constant, $S(t) + E(t) + A(t) + I(t) + R(t) = 1$. 

One trivial solution of the system (\ref{eq:SEAIR_comp_def}) is the disease-free state $S=1$, when the pathogen is absent. If some virus is introduced, an epidemic can occur. Then there is an endemic equilibrium which corresponds to the situation when the fraction of susceptibles is not sufficient for further spread of the disease. When epidemic occurs, the number of unaffected people can be obtained by standard technique which will be applied here \cite{brauer2019mathematical}.  
For the SEAIR model, if we sum the top four equations in (\ref{eq:SEAIR_comp_def}), the following relationship will hold 
\begin{equation}
    \frac{d(S+E+A+I)}{dt} = -\mu(A+I).
\end{equation}
In situations when the epidemic starts with negligibly small number of virus bearers, by taking that at the finish of the epidemic the fractions of individuals with the pathogen is zero, after integration of the last equation, one obtains
\begin{equation}
    S(0) - S(\infty) = \mu \int_0^{\infty} \left[A(t) + I(t)\right]dt. \label{eq:delta_S_through_A_and_I}
\end{equation}

The first equation in (\ref{eq:SEAIR_comp_def}) can be rewritten as 
\begin{equation}
    \frac{dS}{S} = -(\alpha A + \beta I) dt,
\end{equation}
which by integration will result in another relationship between the initial and the final fractions of susceptibles 
\begin{equation}
    \ln\frac{S(0)}{S(\infty)} =\alpha\int_0^{\infty}A(t) dt+  \beta\int_0^{\infty}I(t)dt. \label{eq:log_S_through_A_and_I}
\end{equation}

One can also integrate the fourth equation in (\ref{eq:SEAIR_comp_def}) on both sides to obtain 
\begin{equation}
    I_{\infty} - I_0 = \sigma\int_0^{\infty} A(t) dt - \mu\int_0^{\infty} I(t) dt \approx 0.
\end{equation}
The last result provides a relationship between asymptomatic and infectious fractions in the course of the whole epidemic 
\begin{equation}
    \sigma\int_0^{\infty} A(t) dt = \mu\int_0^{\infty} I(t) dt. \label{eq:total_A_VS_I}
\end{equation}
By combining the relationships (\ref{eq:delta_S_through_A_and_I}), (\ref{eq:log_S_through_A_and_I}) and (\ref{eq:total_A_VS_I}), the following equation for the fraction of unaffected individuals is obtained 
\begin{equation}
  S(0) - S(\infty) = \frac{\mu(\mu + \sigma)}{\alpha\mu + \beta\sigma} \left[\ln S(0) - \ln S(\infty)\right]. \label{eq:final_fract_suscept}
\end{equation}
Using the fact that $f(x) = \ln x$ is steeper than $g(x) = x$ for $x < 1$, one can verify that
\begin{equation}
    S(0) - S(\infty) <  \ln S(0) -\ln S(\infty). \label{eq:dx_vs_dlnx}
\end{equation}
This implies that the transcendental equation (\ref{eq:final_fract_suscept}) has a solution only if
\begin{equation}
    \mu(\mu + \sigma) < \alpha\mu + \beta\sigma. \label{eq:epidemic_emerg_threshold}
\end{equation}
The last inequality is the condition of existence of endemic equilibrium $S=S(\infty); E=A=I=0; R=S(0) - S(\infty)$ of the system (\ref{eq:SEAIR_comp_def}). 

To study the linear stability of the equilibrium states, one should linearize the system (\ref{eq:SEAIR_comp_def}). The respective Jacobian matrix $\mathbf{J} = [ \partial \Dot{B}/\partial C]; B,C \in \{S,E,A,I,R\}$, reads
\begin{equation}
    \mathbf{J}=\begin{bmatrix}
  -\alpha A - \beta I &
    0 &
    -\alpha S & -\beta S & 0 \\[1ex] 
  \alpha A + \beta I &
 -\gamma & \alpha S & \beta S & 0 \\[1ex]
  0 & \gamma & -\sigma-\mu & 0
    & 0 \\[1ex]
    0 & 0 & \sigma & -\mu & 0\\[1ex]
    0 & 0 & \mu & \mu & 0\\
\end{bmatrix}.\label{eq:SEAIR_comp_Jacobian}
\end{equation}
At the disease-free state for which $S=1$ and $E=A=I=R=0$, the Jacobian has rather simple form
\begin{equation}
        \mathbf{J}_{\text{DF}}=\begin{bmatrix}
  0 &
    0 &
    -\alpha & -\beta & 0 \\[1ex] 
  0 &
 -\gamma & \alpha & \beta & 0 \\[1ex]
  0 & \gamma & -\sigma-\mu & 0
    & 0 \\[1ex]
    0 & 0 & \sigma & -\mu & 0\\[1ex]
    0 & 0 & \mu & \mu & 0\\
\end{bmatrix}.\label{eq:SEAIR_comp_Jacobian_simpl}
\end{equation}
Because the first and last columns are zero, this Jacobian has two trivial zero eigenvalues, while the other three are the roots of the polynomial
\begin{equation}
\mathcal{R}(\lambda) = (-\mu-\lambda)\left[(-\gamma - \lambda)(-\sigma-\mu-\lambda) -\alpha\gamma\right] 
    + \beta\gamma\sigma.\label{eq:char_func_compart_main}
\end{equation}
We show in Appendix \ref{A:Stab_compart} that the three nontrivial eigenvalues of the Jacobian have negative real part, which implies linear stability of the disease-free state, if the following relationship holds
\begin{equation}
    \mu(\mu + \sigma) > \alpha\mu + \beta\sigma. \label{eq:Jacob_neg_real}
\end{equation}
The obtained inequality is opposite of the condition for the existence of endemic equilibrium, as one can expect. If (\ref{eq:Jacob_neg_real}) holds, (\ref{eq:epidemic_emerg_threshold}) does not, the disease-free state is stable and epidemic will not occur. In the opposite, if (\ref{eq:Jacob_neg_real}) is not satisfied, the equilibrium $(1, 0, 0, 0, 0)$ is unstable, the epidemic will ensue, and the size of unaffected population can be obtained from (\ref{eq:final_fract_suscept}). Thus the threshold at which epidemic can emerge is the following relationship
\begin{equation}
    \mu(\mu + \sigma) = \alpha\mu + \beta\sigma. \label{eq:Compart_threshold}
\end{equation}

The linear stability analysis of the endemic equilibrium can be applied by the same procedure as for the disease-free one. In this case, in the Jacobian (\ref{eq:SEAIR_comp_Jacobian}) one should take $S = S(\infty)$ and $E = A = I = 0$. By applying the same procedure which is explained in Appendix \ref{A:Stab_compart}, it will be obtained that the only difference from the disease-free case is that instead of $\alpha$ and $\beta$, it should be used $S(\infty)\alpha$ and $S(\infty)\beta$, respectively. This would simply change the condition for stability of the endemic equilibrium to
\begin{equation}
    \mu(\mu + \sigma) > S(\infty)(\alpha\mu + \beta\sigma). \label{eq:Jacob_neg_real_endemic}
\end{equation}
From (\ref{eq:final_fract_suscept}) one has the following relationship for the fraction containing the parameters 
\begin{equation}
      \frac{\mu(\mu + \sigma)}{\alpha\mu + \beta\sigma} = \frac{S(0) - S(\infty)}{\ln S(0) - \ln S(\infty)}.
\end{equation}
Plugging the last relationship in  (\ref{eq:Jacob_neg_real_endemic}), it will be obtained that the endemic equilibrium is stable once the following holds
\begin{equation}
    \frac{S(0) - S(\infty)} {\ln S(0) - \ln S(\infty)} > S(\infty).
\end{equation}
The last inequality can be rearranged as 
\begin{equation}
    \frac{S(0)}{S(\infty)} > 1 + \ln \frac{S(0)}{S(\infty)},\label{eq:stab_condition_compart_enedmic}
\end{equation}
which holds always since $S(0) > S(\infty)$.


\section{Discrete-time SEAIR model}
\label{SEC:Discrete}

Subsequent investigation of a disease spreading model when one needs to account for the contacts between the individuals, is to study epidemic spreading on complex networks framework. Let us consider discrete-time evolution version of the proposed SEAIR model with finite population of $N$ individuals. The network of contacts is conveniently modeled with fixed undirected graph, in which the vertices are the individuals, while the links exist between those persons which have contact to each other. This means that the disease can be transmitted only between neighbors in the graph. An exact approach for analysis of contact-based spreading on complex networks is based on using indicator random variable for each state of every node in the network and working with a system of size $5^N$. However, as it is elaborated in details in \cite{van2014exact}, under the assumption that a node can be in certain state independently of the states of the other nodes, one can instead use the probability of being at that state as a more convenient variable of interest. We proceed in that spirit and build our model on the set of probabilities for each node being in certain state, S, E, A, I, or R, at given moment $n$. We will also denote the parameters with the same Greek letters as in the compartmental model. They correspond to the same transitions and have meaning of probabilities instead of rates. To be more precise, $\alpha$ is the probability that asymptomatic person will infect a susceptible neighbor at one time step, while $\beta$ is the respective probability in the case of contact between infectious and susceptible individual. Once becoming exposed, the person can proceed into asymptomatic phase with probability $\gamma$ at one time step, or remain in the same state with probability $1-\gamma$. The respective probability to show symptoms by asymptomatic individual is $\sigma$. Again, as in the compartmental model, we assume identical probability $\mu$ to become recovered, for both the asymptomatic and the infectious state.

The dynamics of the discrete-time version of the SEAIR model is built similarly to the model considered in \cite{wang2003epidemic}. Denote the probabilities that the individual $i$ at the discrete moment $n$ is in respective state with $p_{S,i}(n)$,  $p_{E,i}(n)$, $p_{A,i}(n)$, $p_{I,i}(n)$  and $p_{R,i}(n)$. Our model assumes reactive process of epidemic spreading, which means that in each time step every individual has contact with every neighbor \cite{gomez2010discrete}. Then, certain susceptible person $i$ at the moment $n+1$ will not receive the infection from any of its neighbors with probability \cite{wang2003epidemic, gomez2010discrete}
\begin{equation}
    \mathcal{P}_i(n) = \prod_{j\in \mathcal{N}_i} \left[1-\alpha p_{A,j}(n)-\beta p_{I,j}(n)\right], \label{eq:Prob_not_infected}
\end{equation}
where $\mathcal{N}_i$ denotes the set of neighbors of $i$. The individual will remain susceptible at the moment $n+1$, if he, or she, did not receive the contagion, which means that
\begin{equation}
    p_{S,i}(n+1) = p_{S,i}(n) \mathcal{P}_i(n). \label{eq:Susc_def}
\end{equation}
Otherwise, an individual can become exposed if he, or she has been susceptible before and received the contagion, or continue to be exposed at the next moment, if the incubation has not finished, with probability 
\begin{equation}
    p_{E,i}(n+1) = p_{S,i}(n)\left[1 - \mathcal{P}_i(n) \right] + (1-\gamma) p_{E,i}(n). \label{eq:Expos_def}
    \end{equation}
The probability of being asymptomatic at the next moment is
\begin{equation}
    p_{A,i}(n+1) = \gamma p_{E,i}(n) + (1 - \sigma-\mu) p_{A,i}(n), \label{eq:Asymp_def}
\end{equation}
where the last term accounts for the situation that neither the symptoms will appear, nor healing will happen in one time step, which imposes a restriction $\mu + \sigma < 1$. The node $i$ will be in state I at the moment $n+1$ with probability
\begin{equation}
    p_{I,i}(n+1) = \sigma p_{A,i}(n) +  (1-\mu)p_{I,i}(n), \label{eq:Infect_def}
\end{equation}
where the former term describes the probability to show symptoms,  if in the previous moment the node was asymptomatic, while the last one corresponds to recovering. Finally, the probability of being recovered at some moment is
\begin{equation}
p_{R,i}(n+1) = p_{R,i}(n) + \mu \left[p_{A,i}(n) + p_{I,i}(n)\right]. \label{eq:Recov_def}
\end{equation}
The set of equations (\ref{eq:Susc_def}) to (\ref{eq:Recov_def}) determines a  discrete-time dynamical system of equations for evolution of probabilities of the states for each node in the network. It can be solved numerically for arbitrary initial condition and one can thus observe the progress of the epidemic at each moment. In practice one can make such studies with networks with size depending on the computational capacities at hand.

\subsection{Epidemic spreading on random regular graphs} 

We will pursue our analysis of spreading processes on random regular graphs where each node has the same degree $k$. For infinitely large random regular graphs, the probabilities of the states are equal for all nodes and one can drop the index of the node. The probability to avoid infection (\ref{eq:Prob_not_infected}) will be simplified to
\begin{equation}
    \mathcal{P}(n) = \left[1-\alpha p_{A}(n)-\beta p_{I}(n)\right]^k. \label{eq:Prob_not_infec_simple}
\end{equation}
Then the system of equations for discrete-time epidemic spreading on random regular graph reads
\begin{eqnarray}
    p_{S}(n+1) &=& p_{S}(n) \left[1-\alpha p_{A}(n)-\beta p_{I}(n)\right]^k,\nonumber\\
    p_{E}(n+1) &=& p_{S}(n) \left\{1 -\left[1-\alpha p_{A}(n) - \beta p_{I}(n)\right]^k \right\}\nonumber\\ &+& (1-\gamma) p_{E}(n),\nonumber\\
    p_{A}(n+1) &=& \gamma p_{E}(n) + (1-\sigma-\mu) p_{A}(n), \nonumber\\
    p_{I}(n+1) &=& \sigma p_{A}(n) +  (1-\mu)p_{I}(n),\nonumber\\
    p_{R}(n+1) &=& p_{R}(n) + \mu \left[p_{A}(n) + p_{I}(n)\right]. \label{eq:SEAIR_random_reg}
\end{eqnarray} 

Here we have also two equilibrium points: one where all individuals are susceptible $\mathbf{p} = (1, 0, 0, 0, 0)$ and the other when the fraction of susceptible individuals is such that it prevents further spread of the disease $\mathbf{p} = (p_S^*, 0, 0, 0, 1-p_S^*)$. In the Appendix \ref{A:Rand_reg_endemic} we show how application of similar reasoning as for the compartmental model can allow to find a closed form equation for the number of susceptible individuals at the end of the epidemic. It can be applied when the contagion probabilities $\alpha$ and $\beta$ are very small. 
As it is shown in the Appendix \ref{A:Rand_reg_endemic}, the equation for determination of the probability of the susceptible state at the end of epidemic is very similar to the respective one for the compartmental case 
\begin{equation}
  p_{S}(0) - p_{S}(\infty) = \frac{\mu(\mu + \sigma)}{k(\alpha\mu + \beta\sigma)} \ln\frac{p_{S}(0)}{p_{S}(\infty)}. \label{eq:susc_at_end_of_Epidemic}
\end{equation}
The last result extends the one for all-to-all coupling considered in compartmental models, where effectively each individual can get the disease from anyone in the population. Here, it is obtained that appropriately modified relationship holds for restricted number of contacts. By repeating the same analysis as for the compartmental model, one can also obtain that the condition for existence of endemic equilibrium is
\begin{equation}
    \mu(\mu + \sigma) < k(\alpha\mu + \beta\sigma). \label{eq:epidemic_emerg_thresh_rand_reg}
\end{equation}
One can note that the last relationship is similar to the respective one for the compartmental model (\ref{eq:epidemic_emerg_threshold}), and the only difference is the presence of the node degree $k$ in the discrete-time case.

We can further make linear stability analysis of equilibria by linearizing the evolution equations. The respective Jacobian matrix at the fixed points for which $p_E = p_A = p_I = 0$ and $p_S = p_S^*$ is
\begin{equation}
    \mathbf{J} =
\begin{bmatrix}
  1 &
    0 &
    -k\alpha p_S^* & -k\beta  p_S^* & 0 \\[1ex] 
  0 &
 1 - \gamma & k\alpha p_S^* & k\beta p_S^* & 0 \\[1ex]
  0 & \gamma & 1 - \sigma -\mu & 0
    & 0 \\[1ex]
    0 & 0 & \sigma & 1 - \mu & 0\\[1ex]
    0 & 0 & \mu & \mu & 1\\
\end{bmatrix}. \label{eq:SEAIR_Jacobian_rand_reg}
\end{equation}
Solving the characteristic equation $\mathcal{Q}(\lambda) = \det (\mathbf{J} - \lambda \mathbf{I}) = 0$ of the Jacobian (\ref{eq:SEAIR_Jacobian_rand_reg}), for $p_S^* = 1$ will result in two trivial eigenvalues equal to one and other three. We note that in the discrete-time case the trivial eigenvalues have value equal to one that also corresponds to the marginal stability, which are zero in the continuous-time scenario. The nontrivial eigenvalues $\lambda$ are the roots of the polynomial
\begin{eqnarray}
     \mathcal{S}(\lambda) &=& (1-\mu-\lambda)
 \left[(1-\gamma-\lambda)(1-\sigma-\mu-\lambda) - \alpha\gamma k \right]\nonumber\\ &+&\beta\gamma\sigma k.\label{eq:SEAIR_determinant_main}
\end{eqnarray}
One can notice the similarity to the respective characteristic polynomial for the compartmental case (\ref{eq:char_func_compart_main}). The only difference is presence of the degree $k$ which in the latter expression multiplies $\alpha$ and $\beta$ and one has $1-\lambda$ in the discrete-time case instead of $-\lambda$. In the Appendix \ref{A:Stab_rand_reg} it is shown that the disease-free state is stable once the following relationship is satisfied 
\begin{equation}
    \mu(\mu+\sigma) > k(\alpha \mu +\beta\sigma), \label{eq:Dis-free_stab_rand_reg}
\end{equation}
which is similar to the one for the compartmental case. Again, the only difference is the presence of the node degree $k$. 

The linear stability of the endemic equilibrium is established from the leading eigenvalue of the same Jacobian matrix (\ref{eq:SEAIR_Jacobian_rand_reg}) as the disease-free one, but using $p_S^*=p_{S}(\infty)$ obtained from (\ref{eq:susc_at_end_of_Epidemic}). The procedure is nearly the same as for the disease-free equilibrium and the only difference is that instead of $k$ one should use $k p_{S}(\infty)$ in all analysis. Then the endemic equilibrium will be stable, if the condition similar to (\ref{eq:Dis-free_stab_rand_reg}) is satisfied
\begin{equation}
    \mu(\mu+\sigma) > k p_{S}(\infty)(\alpha \mu +\beta\sigma). \label{eq:Endem_stab_rand_reg}
\end{equation}
Without showing the details, we will just mention that once the endemic equilibrium in this discrete-time disease spreading model exists, it is linearly stable.

\subsection{Epidemic spreading on complex networks} 

Let us now consider the general case when the contacts between individuals are  described with complex network. In studies of interacting units coupled in a network it is typical to define the state of the whole system by stacking the state vectors of each node one on top of another. In this case another ordering is more appropriate \cite{prakash2010got}. First, create vector of the probabilities of susceptible states of all nodes $\mathbf{p}_S = [p_{S,1}, p_{S,2}, \dots, p_{S,N}]^{\text{T}}$, then those of the exposed states $\mathbf{p}_E = [p_{E,1}, p_{E,2}, \dots, p_{E,N}]^{\text{T}}$, and likewise for the remaining three $\mathbf{p}_A$, $\mathbf{p}_I$ and $\mathbf{p}_R$. Also, denote with $\mathbf{A}$ the adjacency matrix of the network of contacts between the individuals with elements $A_{i,j} = 1$ only if nodes $i$ and $j$ are neighbors and $A_{i,j} = 0$ otherwise. Under general circumstances, determination of the probabilities at the end of epidemic in case it happens, is very complicated, if not impossible. However, when the contagiousness is weak, which means that $\alpha \ll 1$ and $\beta \ll 1$, one can obtain similar expressions which relate initial and final probabilities of susceptible state as for the former two models. As it is explained in details in 
the Appendix \ref{A:Complex_endemic}, when $\alpha \ll 1$ and $\beta \ll 1$, the susceptibility probability vector can be calculated from the following self-consistent system
\begin{eqnarray}
\mathbf{p}_{S}(0) - \mathbf{p}_{S}(\infty) &=& \mu (1 + \frac{\sigma}{\mu}) \sum_{n=0}^{\infty}  \mathbf{p}_{A}(n),\nonumber\\
    \ln \mathbf{p}_{S}(0) - \ln \mathbf{p}_{S}(\infty) &=& \left(\alpha + \beta\frac{\sigma}{\mu}\right) \mathbf{A} \sum_{n=0}^{\infty}  \mathbf{p}_{A}(n). \label{eq:Susc_endemic_complex}
\end{eqnarray}
The solution of the system of transcendental equations (\ref{eq:Susc_endemic_complex}) consists of the susceptibility vector at the endemic equilibrium $\mathbf{p}_{S}(\infty)$ and the vector of sums of probabilities of asymptomatic states
 \begin{equation}
     \mathbf{p}_A = \sum_{n=0}^{\infty}  \mathbf{p}_{A}(n). 
     \label{eq:sums_p_A}
 \end{equation}
Such transcendental system should be solved numerically, and for large networks might be impossible task. However, one can at least obtain  how the solution will look like, when the epidemic is weak in a sense that only small fraction of the population is infected during its course. In such case the probability of susceptibility will not change significantly $p_{S}(0) \approx p_{S}(\infty)$. This situation might be present, for example, when the contagiousness of the pathogens is slightly over the threshold. Then one can keep only the leading terms in the expansion of the logarithm and obtain 
\begin{equation}
    \ln p_{S,i}(0) - \ln p_{S,i}(\infty) \approx p_{S,i}(0) - p_{S,i}(\infty).
\end{equation}
The last approximation means that effectively the left hand sides of relationships (\ref{eq:Susc_endemic_complex}) are equal. Then, after some algebra, by using (\ref{eq:sums_p_A}), from those relationships one can obtain
\begin{equation}
    \mathbf{p}_A = \frac{\alpha\mu + \beta\sigma}{\mu(\mu + \sigma)} \mathbf{A} \mathbf{p}_A. \label{eq:sum_pA_self_consist}
\end{equation}
The last relationship is eigenvalue equation of a matrix which is the adjacency matrix multiplied by the scalar $(\alpha\mu + \beta\sigma)/[\mu(\mu + \sigma)]$, which corresponds to eigenvalue equal to one. Thus, the vector of sums of the probabilities of the asymptomatic state (\ref{eq:sums_p_A}) represents eigenvector of the adjacency matrix that corresponds to the eigenvalue $\Lambda$ such that
\begin{equation}
    1 = \Lambda\frac{\alpha\mu + \beta\sigma}{\mu(\mu + \sigma)}. \label{eq:L_adjacency_condition}
\end{equation}
To determine which is the eigenvalue $\Lambda$, observe that
we can apply similar inequality as (\ref{eq:dx_vs_dlnx}), which means that for each node $i$ one has
\begin{equation}
    p_{S,i}(0) - p_{S,i}(\infty) < \ln p_{S,i}(0) - \ln p_{S,i}(\infty).
\end{equation}
This implies that one has the following vector inequality
\begin{equation}
   \mathbf{p}_{A} < \frac{\alpha\mu + \beta\sigma}{\mu(\mu + \sigma)} \mathbf{A}\mathbf{p}_{A}, \label{eq:asymp_ineq_comp_net}
\end{equation}
which is obtained from (\ref{eq:Susc_endemic_complex}) with simple algebra. When the epidemic parameters are such that 
\begin{equation}
     \Lambda_{\max}\frac{\alpha\mu + \beta\sigma}{\mu(\mu + \sigma)} < 1, \label{eq:L_max_adj_ineq}
\end{equation}
where $\Lambda_{\max}$ is the largest eigenvalue of $\mathbf{A}$, then (\ref{eq:L_adjacency_condition}) can not be satisfied for no one eigenvalue. That is the condition when endemic equilibrium does not exist. To determine when it will emerge, one should increase the value of the fraction in (\ref{eq:L_max_adj_ineq}), by modifying the epidemic parameters. Then, the first eigenvalue that can satisfy the equation as (\ref{eq:L_adjacency_condition}) will be exactly the largest eigenvalue $\Lambda_{\max}$. Thus the condition for existence of endemic equilibrium is
\begin{equation}
    \mu(\mu + \sigma) < \Lambda_{\max}(\alpha\mu + \beta\sigma). \label{eq:Condit_existence_endemic_complex}
\end{equation}
The last result is generalization of the case of random regular graph for which $\Lambda_{\max} = k$. 

Because the leading eigenvalue of the adjacency matrix $\mathbf{A}$ determines the equation for $\mathbf{p}_A$, that vector is determined from the respective eigenvector, or the principal eigenvector of $\mathbf{A}$.
However, the eigenvalue equation (\ref{eq:sum_pA_self_consist}) just determines the relative magnitudes of the components of $\mathbf{p}_A$. If it is used in the system (\ref{eq:Susc_endemic_complex}), it will be obtained that for each node the change in the probability of being susceptible is the same, which counters the fact that it corresponds to the respective component of the principal eigenvector. We further note that from (\ref{eq:Susc_endemic_complex}) the relative magnitudes of the changes of the probabilities of susceptible state $\mathbf{p}_{S,i}(0) - \mathbf{p}_{S,i}(\infty)$, and the resulting probabilities of recovered state $\mathbf{p}_{R,i}(\infty)$, as collinear to $\mathbf{p}_A$, are also proportional to the principal eigenvector of the adjacency matrix. This is in accordance to the reasoning that the individuals with highest risk of infection are those with many contacts, and particularly those which have many high-degree neighbors. Thus the eigenvector centrality of the node \cite{bonacich1972factoring}, which is the respective component in the principal eigenvector, determines the risk of infection of that node. We note that, by applying this procedure one can also show that the same conclusions about the role of the leading eigenvalue and principal eigenvector in epidemic spreading on complex networks hold for the simpler SEIR, SIR and SIS models. 

It should be emphasized that, there are works in the literature which point that the principal eigenvector of the adjacency matrix can have important role in disease spreading. The probability of infectious state of a node in SIS spreading model on complex networks was found to be proportional to the eigenvector centrality, in vicinity of the epidemic threshold \cite{goltsev2012localization}, similarly as the analysis above claims. This finding was further extended to multilayer networks \cite{de2017disease}. It was reported in the same contributions, that if the structure of the network is such that the principal eigenvector is localized, the spreading will be limited to small number of nodes, even for large networks. Thus, the association between the eigenvector centrality and probability of becoming infected might not be observed in certain scenarios. Such example could be observed, for example, in structured network with communities, when the virus starts spreading in a node in one community, while principal eigenvector has significantly large components in the other communities. Thus, more research on this issue is needed for better understanding of the conditions when the principal eigenvector is really useful in estimation of the risk of infection. Finally, it is worth noting that it was found that the principal eigenvector of another matrix -- the submatrix corresponding to the infectious states -- also determines the disease spreading pathways. This observation has appeared in the studies of disease spreading between spatial regions in a waterborne disease \cite{gatto2012generalized} and COVID-19 \cite{gatto2020spread}. 

We now proceed with the study of the stability of the disease-free equilibrium and determine the epidemic threshold. The associated Jacobian matrix is obtained by taking the respective derivatives in the equations (\ref{eq:Susc_def}) to (\ref{eq:Recov_def}).  Also, we remind that after making differentiation, at the epidemic inception, in the Jacobian it should be taken $p_{E,i}=0$, $p_{A,i}=0$, $p_{I,i}=0$ and $p_{S,i}=1$. It can be verified that the Jacobian will have the following matrix form
\begin{equation}
       \mathbf{J} =
\begin{bmatrix}
  \mathbf{I} &
    \mathbf{0} &
    -\alpha \mathbf{A} & -\beta \mathbf{A} & \mathbf{0} \\[1ex] 
  \mathbf{0} &
 (1-\gamma)\mathbf{I} & \alpha \mathbf{A} \ & \beta \mathbf{A} & \mathbf{0} \\[1ex]
  \mathbf{0} & \gamma \mathbf{I} & (1 - \sigma -\mu) \mathbf{I} & \mathbf{0}
    & \mathbf{0} \\[1ex]
    \mathbf{0} & \mathbf{0} & \sigma\mathbf{I} & (1 - \mu) \mathbf{I} & \mathbf{0}\\[1ex]
    \mathbf{0} & \mathbf{0} & \mu \mathbf{I}& \mu \mathbf{I}& \mathbf{I}\\
\end{bmatrix}, \label{eq:SEAIR_Jacobian_complex_dis-free}
\end{equation}
where $\mathbf{I}$ is identity matrix of the same size $N$ as the adjacency matrix of the network $\mathbf{A}$ -- the number of nodes in the network. One can note the similarity in the structure between the last matrix and that in (\ref{eq:SEAIR_Jacobian_rand_reg}). The eigenvalues of the last Jacobian are obtained from the characteristic equation $\mathcal{T}(\lambda) = \det(\mathbf{J} - \lambda \mathbf{I}_{5N}) = 0$, where we emphasize that the involved identity matrix has size $5N\times 5N$. One could use the approach given in \cite{prakash2010got} to determine the dependence of the epidemic threshold on the largest eigenvalue of the adjacency matrix. We have chosen alternative approach here, based on Schur's determinant identity
\begin{equation}
    \det
\begin{bmatrix}
  \mathbf{Q} &
    \mathbf{R}\\[1ex] 
  \mathbf{S} &
 \mathbf{T} \end{bmatrix} = \det(\mathbf{T})\cdot \det(\mathbf{Q} - \mathbf{R}\mathbf{T}^{-1}\mathbf{S}), \label{eq:Schur's_det_identity}
\end{equation}
which is more general. Clearly, when a matrix has many zero submatrices, its application provides simpler results. By repetitive use of it, which is elaborated in the Appendix \ref{A:SEAIR_Charact_polynomial}, it can be shown that the nontrivial eigenvalues can be obtained from the polynomial corresponding to the following determinant
\begin{equation}
   \mathcal{V}(\lambda) = \det\left[\frac{(1-\gamma-\lambda)(1-\sigma-\mu-\lambda)(1-\mu-\lambda)}{\gamma[\alpha(1-\mu-\lambda) - \beta\sigma]}\mathbf{I} - \mathbf{A} \right].
\end{equation}
We note that the same determinant can be obtained by the procedure given in the Appendix \ref{A:eigenval_eignvec_Jacob} which even delivers the eigenvectors of the Jacobian. Currently, we cannot state whether the approach in Appendix \ref{A:SEAIR_Charact_polynomial} is just an alternative, or it might have potential to provide results when the latter is not useful. 
To continue with the analysis, one can substitute the multiplier of the identity matrix in the last equation as
\begin{equation}
    \Lambda = \frac{(1-\gamma-\lambda)(1-\sigma-\mu-\lambda)(1-\mu-\lambda)}{\gamma[\alpha(1-\mu-\lambda) - \beta\sigma]},\label{eq:Two_lambdas_main}
    \end{equation}
and will obtain the characteristic function for determination of the eigenvalues of the adjacency matrix, $\det (\Lambda \mathbf{I} -  \mathbf{A})$. From the relationship (\ref{eq:Two_lambdas_main}) it can be seen that to each eigenvalue of the adjacency matrix $\Lambda$ correspond three eigenvalues of the Jacobian $\lambda$, which are obtained from the polynomial
\begin{eqnarray}
    \mathcal{S}_{\Lambda}(\lambda) &=&  (1-\mu-\lambda)\left[(1-\gamma-\lambda)(1-\sigma-\mu-\lambda) - \alpha\gamma\Lambda  \right] \nonumber \\ &+& \beta\gamma\sigma\Lambda. \label{eq:lambda_poly_complex}
\end{eqnarray}
The last relationship will be identical to the respective one for the random regular graph (\ref{eq:SEAIR_determinant_main}), if one substitutes $\Lambda$ with $k$. Because all eigenvalues $\Lambda$ of the  adjacency matrix are real \cite{boccaletti2006complex}, the coefficients in the last polynomial in $\lambda$ are also real for each $\Lambda$. Instead of checking whether the roots of the last polynomial are within the unit circle, one can  determine the eigenvalues of the related polynomial that corresponds to the Jacobian of the compartmental model (\ref{eq:char_func_compart_main}), and then use the relationship between the roots of the characteristic polynomials of the compartmental and discrete-time models (\ref{eq:_poly_roots_relation}). The disease-free state of the model on complex network is unstable, if there is at least one eigenvalue $\Lambda$ of the adjacency matrix, for which there is a real positive root of the polynomial (\ref{eq:lambda_poly_complex}) in which one should use $-\lambda$ instead of $1-\lambda$. This situation happens when parameters are such that the following inequality holds
\begin{equation}
    \mu(\mu + \sigma) < \Lambda(\alpha\mu + \beta\sigma).
\end{equation}
In determination of the epidemic threshold in SIS and SIR models, usually the fraction between the contagiousness and recovery parameters is varied. Since the last inequality is a bit more complex than those in the SIS and SIR models, for the SEAIR model one can use the parameters $\alpha$ and $\beta$ as bifurcation parameters. Then from the last inequality the smallest $\alpha$ and $\beta$ for which the disease-free state is unstable are those obtained for the largest $\Lambda$, that is the leading eigenvalue of the adjacency matrix $\Lambda_{\max}$. Thus, the same condition (\ref{eq:Condit_existence_endemic_complex}) implies existence of endemic equilibrium and instability of the disease-free state. If the opposite is true,
\begin{equation}
    \mu(\mu + \sigma) > \Lambda_{\max} (\alpha \mu + \beta\sigma),
    \label{eq:stab_dis-free_complex}
\end{equation}
then the disease-free state will be stable. The last inequality can be seen as generalization of the respective one for the random regular graphs (\ref{eq:Dis-free_stab_rand_reg}).


Let us proceed with determination of the linear stability of the endemic state. To determine the respective Jacobian matrix, first observe the following derivatives
\begin{align}
    \frac{\partial p_{S,i}}{\partial p_{A,j}} &= -\alpha p_{S,i} A_{i,j}, &
 &\frac{\partial p_{S,i}}{\partial p_{I,j}} = -\beta p_{S,i} A_{i,j},\nonumber\\
    \frac{\partial p_{E,i}}{\partial p_{A,j}} &= \alpha p_{S,i} A_{i,j}, &
    &\frac{\partial p_{E,i}}{\partial p_{I,j}} = \beta p_{S,i} A_{i,j}, \label{eq:partial_der_endem_complex}
\end{align}
where $A_{i,j}$ is the $i,j$-th element of the adjacency matrix. The remaining partial derivatives in the Jacobian matrix are the same as for the disease-free state and are conveniently captured in the respective submatrices in (\ref{eq:SEAIR_Jacobian_complex_dis-free}). The form of the partial derivatives (\ref{eq:partial_der_endem_complex}) is such that for each $p_{S, i}$ they have identical form and likewise for the $p_{E,i}$. To write a more compact form for expressing such relationship, one can introduce a diagonal matrix $\mathbf{\Sigma}$ which contains the endemic equilibrium probabilities $p_{S,i}(\infty)$ along the diagonal $\Sigma_{i,i} =  p_{S,i}(\infty)$.   
Then, one can obtain that the partial derivatives between the susceptibility and exposed vectors with respect to the asymptomatic and infectious vectors can be compactly written as
\begin{align}
    \frac{\partial \mathbf{p}_{S}}{\partial \mathbf{p}_{A}} &= -\alpha \mathbf{\Sigma A}, &
    &\frac{\partial \mathbf{p}_{S}}{\partial \mathbf{p}_{I}} = -\beta \mathbf{\Sigma A}, \nonumber\\
    \frac{\partial \mathbf{p}_{E}}{\partial \mathbf{p}_{A}} &= \alpha \mathbf{\Sigma A}, &
    &\frac{\partial \mathbf{p}_{E}}{\partial \mathbf{p}_{I}} = \beta \mathbf{\Sigma A}.
\end{align}
Now, the Jacobian of the endemic equilibrium differs from that for the disease-free one, only in that it contains the matrix product $\mathbf{\Sigma A}$ instead of $\mathbf{A}$. Respectively, the stability of the endemic equilibrium will depend on the leading eigenvalue $L_{\max}$ of the matrix product $\mathbf{\Sigma A}$. Thus, the stability condition is similar to that for the disease-free state (\ref{eq:stab_dis-free_complex}) 
\begin{equation}
\mu(\mu + \sigma) > L_{\max}(\alpha\mu + \beta\sigma). \label{eq:endem_stab_complex_main}
\end{equation}
In the Appendix \ref{A:Endem_stability} it is shown that in case of small contagiousness $\alpha \ll 1$ and $\beta \ll 1$ and when epidemic affects small population during its course, the endemic equilibrium is linearly stable.

Let us finally examine the behavior of the disease spreading in the early phase of epidemic. The one-step evolution of the probabilities of different states in disease spreading on complex networks is given by equations (\ref{eq:Susc_def}) to (\ref{eq:Recov_def}). One can combine all probabilities in single column vector as $\mathbf{p} = [\mathbf{p}_S^{\text{T}}, \mathbf{p}_E^{\text{T}}, \mathbf{p}_A^{\text{T}}, \mathbf{p}_I^{\text{T}}, \mathbf{p}_R^{\text{T}}]^{\text{T}}$ and the right hand sides of the probability evolution equations in a vector $\mathcal{F}$. Then, one-step evolution of the probabilities can be compactly written in vector notation as
\begin{equation}
    \mathbf{p}(n+1) = \mathcal{F}\left[\mathbf{p}(n)\right]. \label{eq:Near_dis-free_evol}
\end{equation}
Consider early stages of the epidemics, when the states $\mathbf{p}(n)$ are sufficiently close to the disease-free equilibrium $\mathbf{p}_{\text{DF}} = \mathcal{F}\left[\mathbf{p}_{\text{DF}}\right]$. Denote with $\delta \mathbf{p}(n) = \mathbf{p}(n) - \mathbf{p}_{\text{DF}}$ the deviation from the disease-free state. Then from (\ref{eq:Near_dis-free_evol}) one has
\begin{equation}
    \delta\mathbf{p}(n+1) = \mathbf{p}(n+1) - \mathbf{p}_{\text{DF}}  = \mathcal{F}\left[\mathbf{p}(n)\right] - \mathbf{p}_{\text{DF}}.
\end{equation}
The linear approximation of the nonlinear function $\mathcal{F}$ in vicinity of the disease-free state is
\begin{eqnarray}
    \mathcal{F}\left[\mathbf{p}(n)\right] &\approx& \mathcal{F}\left[\mathbf{p}_{\text{DF}}\right] + \mathbf{J}_{\text{DF}} \left[\mathbf{p}(n) - \mathbf{p}_{\text{DF}}\right] \nonumber \\ &=& \mathbf{p}_{\text{DF}} + \mathbf{J}_{\text{DF}} \delta\mathbf{p}(n), \label{eq:nonlin_evol_linerized}
\end{eqnarray}
where $\mathbf{J}_{\text{DF}}$ is the Jacobian at the disease-free state. It means that consecutive perturbations satisfy simple relationship
\begin{equation}
    \delta\mathbf{p}(n+1) \approx  \mathbf{J}_{\text{DF}} \delta \mathbf{p}(n). \label{eq:evol_of_perturb}
\end{equation}
Thus, at the early phase of an epidemic, the perturbation at given moment $n$ is approximately given as
\begin{equation}
    \delta\mathbf{p}(n) \approx \mathbf{J}_{\text{DF}}^n \delta\mathbf{p}(0).
\end{equation}

Denote with $\mathbf{z}_i$ the eigenvector of the Jacobian $\mathbf{J}_{\text{DF}}$ that corresponds to the eigenvalue $\lambda_i$. Consider situation when the eigenvectors $\mathbf{z}_i$ constitute an orthonormal basis, in which the perturbation $\delta \mathbf{p}(0)$ can be expressed in terms of the Jacobian basis vectors as \begin{equation}
    \delta\mathbf{p}(0) = \sum_{i=1}^{5N} p_i \mathbf{z}_i.
\end{equation}
Then, after $n$ time steps the perturbation will approximately evolve to
\begin{equation}
    \delta\mathbf{p}(n) = \mathbf{J}_{\text{DF}}^n\delta\mathbf{p}(0) = \sum_{i=1}^{5N} p_i  \lambda_i^n \mathbf{z}_i.
\end{equation}
It is clear that as the number of steps $n$ increases, the projection along the principal eigenvector will dominate the others. It means that one can use the approximation
\begin{equation}
    \delta\mathbf{p}(n) \approx p_{\max}  \lambda_{\max}^n \mathbf{z}_{\max},
\end{equation}
where $p_{\max}$ is the projection of the initial perturbation along the principal eigenvector $\mathbf{z}_{\max}$. In the Appendix \ref{A:eigenval_eignvec_Jacob} it is explained that when the epidemic starts the leading eigenvalue of the Jacobian $\lambda_{\max}$ depends on that of the adjacency matrix $\Lambda_{\max}$ and that the principal eigenvector of the Jacobian in that case is determined with the principal eigenvector of the adjacency matrix. Thus, the latter determines the evolution of the epidemics at the early stages. We emphasize that this is an approximation since as $n$ grows, the Jacobian which is used, represents the nonlinear evolution less accurately, because the state of the system goes away from the disease-free one. Although being an approximation, the last result provides an estimate of the risk of being infected of the nodes in a network, by the respective eigenvector centrality.

\section{Numerical experiments and discussion}\label{SEC:Numerical}


The focus of numerical experiments in this work is put on validation of the theoretical results for the discrete-time SEAIR model. The theoretical analysis of the compartmental model was classical and did not bring any significant novelty, which is not known for the other compartmental models. Thus, the potential of the compartmental model should be tested on real data, which is left for future study.

\begin{figure}[t]
\includegraphics[width=0.9\linewidth]{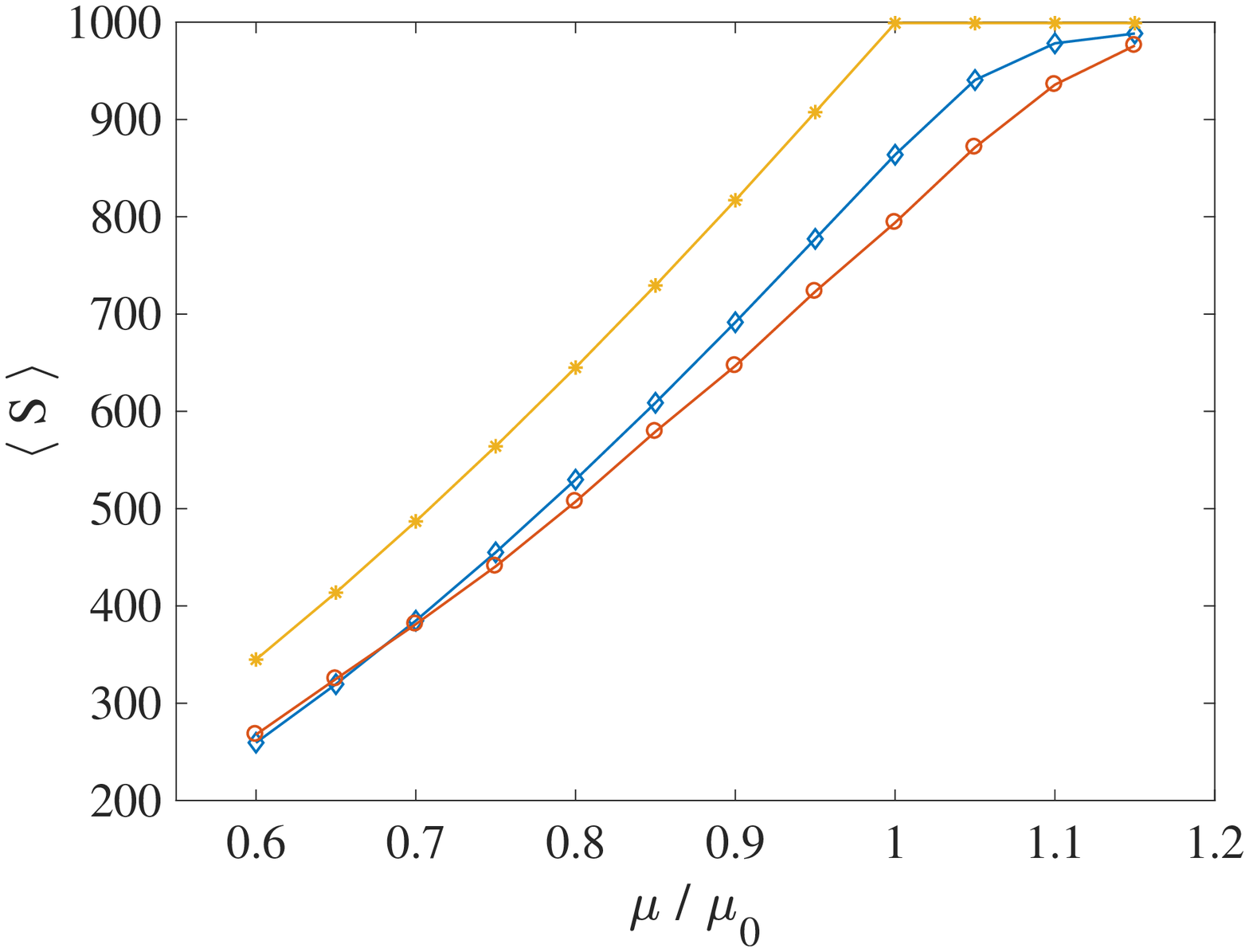}
\caption{\textbf{Disease spreading on random regular graph and random graph with constant degree distribution.}The curves represent the dependence of the number of susceptible individuals at the end of the epidemic
on the parameter $\mu$. The meaning of the symbols is the following: orange stars – theoretical values from eq.  for (\ref{eq:susc_at_end_of_Epidemic})
for infinite-size random regular graph with node degree 50; blue diamonds -- random regular graph with the same degree and 1000 nodes; red circles -- random graph with uniform degree distribution in [30,70] and 1000 nodes. 
\label{fig:Rand_reg}}
\end{figure}

We have made simulations of disease spreading on random regular graphs by numerical solution of the evolution equations (\ref{eq:SEAIR_random_reg}). The aim of these numerical experiments was to check the validity of the epidemic threshold relationships, as well as the equation for the fraction of the susceptible individuals at the end of the epidemic (\ref{eq:susc_at_end_of_Epidemic}). For computational reasons, for this and the other numerical experiments in this work, we have considered networks with 1000 nodes. All versions of the SEAIR epidemic spreading model considered here, have five parameters $\alpha, \beta, \sigma, \mu$ and $\gamma$. However, theoretical analysis in previous sections has shown that only the first four of them are relevant for determination of the epidemic threshold and the susceptible fraction at the end of the epidemic. 
The contagiousness parameters' values, $\alpha = 0.0025$ and $\beta = 0.002 < \alpha$ were chosen arbitrarily, by caring to be small to ensure that the approximations made in the theoretical analysis are justified and using the observation that for COVID-19 contagiousness is bigger before the onset of symptoms. We have taken $\gamma = 0.5$ which should correspond to two days mean period of incubation, while the value of $\sigma = 0.2$ was chosen arbitrarily. The critical parameter $\mu_0$ was calculated from the
following quadratic equation 
\begin{equation}
    k\frac{\alpha\mu_0 + \beta\sigma}{\mu_0(\mu_0 + \sigma)} = 1, \label{eq:mu_quadratic_eq_rand_reg}
\end{equation}
which is obtained from the condition for emergence of endemic state for the random regular graphs (\ref{eq:epidemic_emerg_thresh_rand_reg}). The value of the parameter $\mu$ was varied in the vicinity of $\mu_0$. All simulations were repeated for ten different networks and for each network ten different initializations were made by putting a randomly chosen node in exposed state (patient zero), while leaving the remaining ones as susceptible. The pathogen was considered as extinct at the moment when the total fraction of exposed, asymptomatic and infectious individuals is smaller than $10^{-8}$ of the population. In the figure \ref{fig:Rand_reg} are shown the average number of the susceptible individuals at the end of the epidemic. The number of susceptible individuals for each particular simulation is simply sum of the probabilities of the susceptible state over all nodes. The averaging was performed for all networks from the same type and for all initial conditions. In the blue diamonds are given the results for the random regular graph with node degree $k = 50$, while with red circles are those for random graph with constant degree distribution in the interval $[30, 70]$. We emphasize that in this figure the threshold value $\mu_0$ is obtained for the random regular graph and the same value is used for the others. Both considered kinds of graphs have the same average degree, and thus show similar results, particularly when one is far enough from the threshold $\mu = \mu_0$. In vicinity of $\mu_0$, as was theoretically shown for general complex networks, the epidemic threshold depends on the leading eigenvalue as is given in (\ref{eq:Condit_existence_endemic_complex}), which for the graph with distributed node degree is greater than the average degree $\Lambda_{\max} > \langle k \rangle$ as the Perron-Frobenius theorem claims \cite{boccaletti2006complex}. Thus, for the same $\mu$ one expects more infected individuals for the graphs with distributed degree. The results from the simulations are further compared with theoretical values obtained from (\ref{eq:susc_at_end_of_Epidemic}) for random regular graph with $k=50$ nodes, which holds for network with infinite number of nodes. It can be seen a noticeable difference between the theoretical curve and the simulations. One reason for such discrepancy could be attributed in the fact that the theoretical results hold for infinitely large networks. The other factor is the way of initialization of the epidemics, which even in case of stable disease-free state, $\mu > \mu_0$, produces a small fraction of potentially infected individuals -- at least the neighbors of the patient zero.

\begin{figure*}[t]
\includegraphics[width=0.45\linewidth]{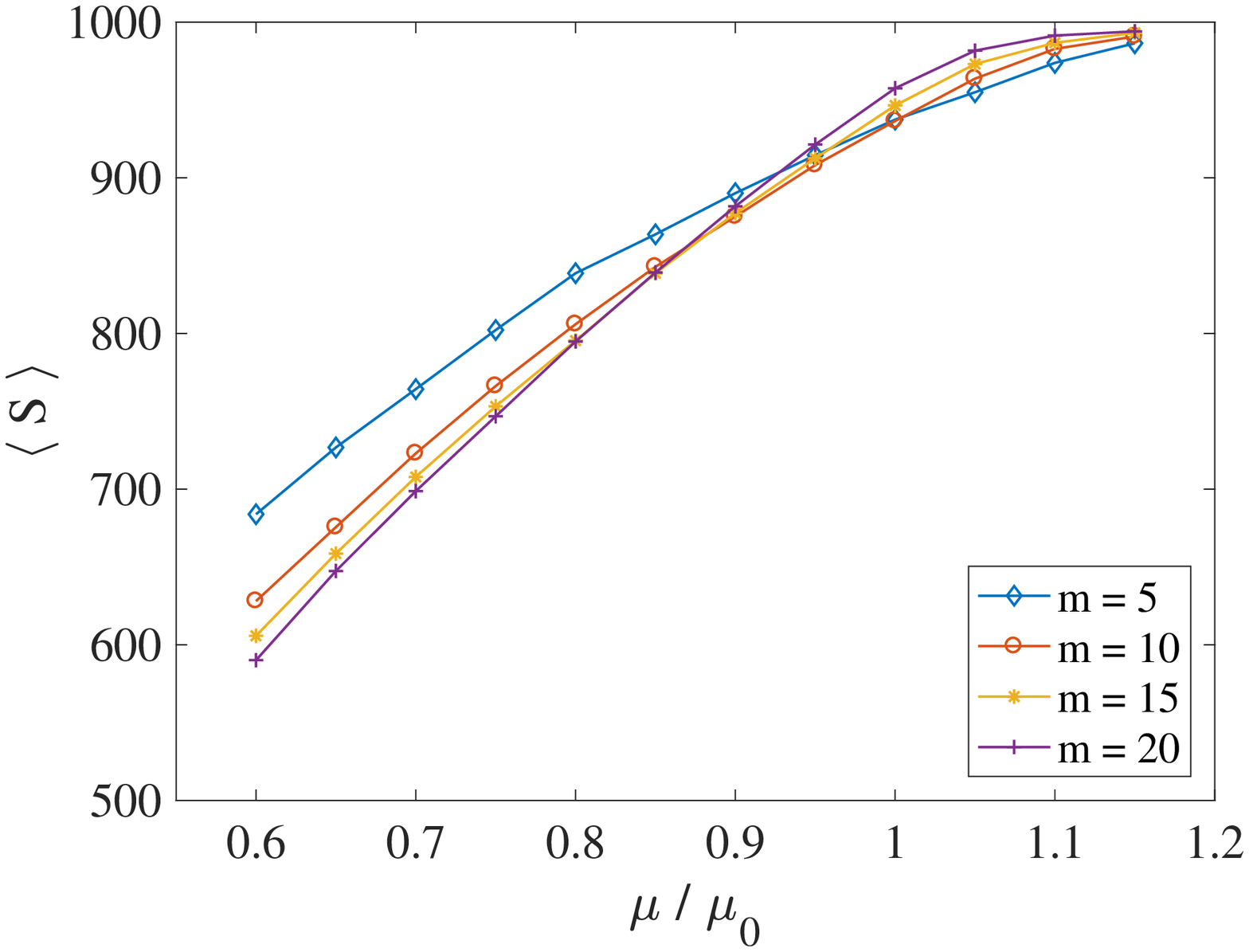}
\includegraphics[width=0.45\linewidth]{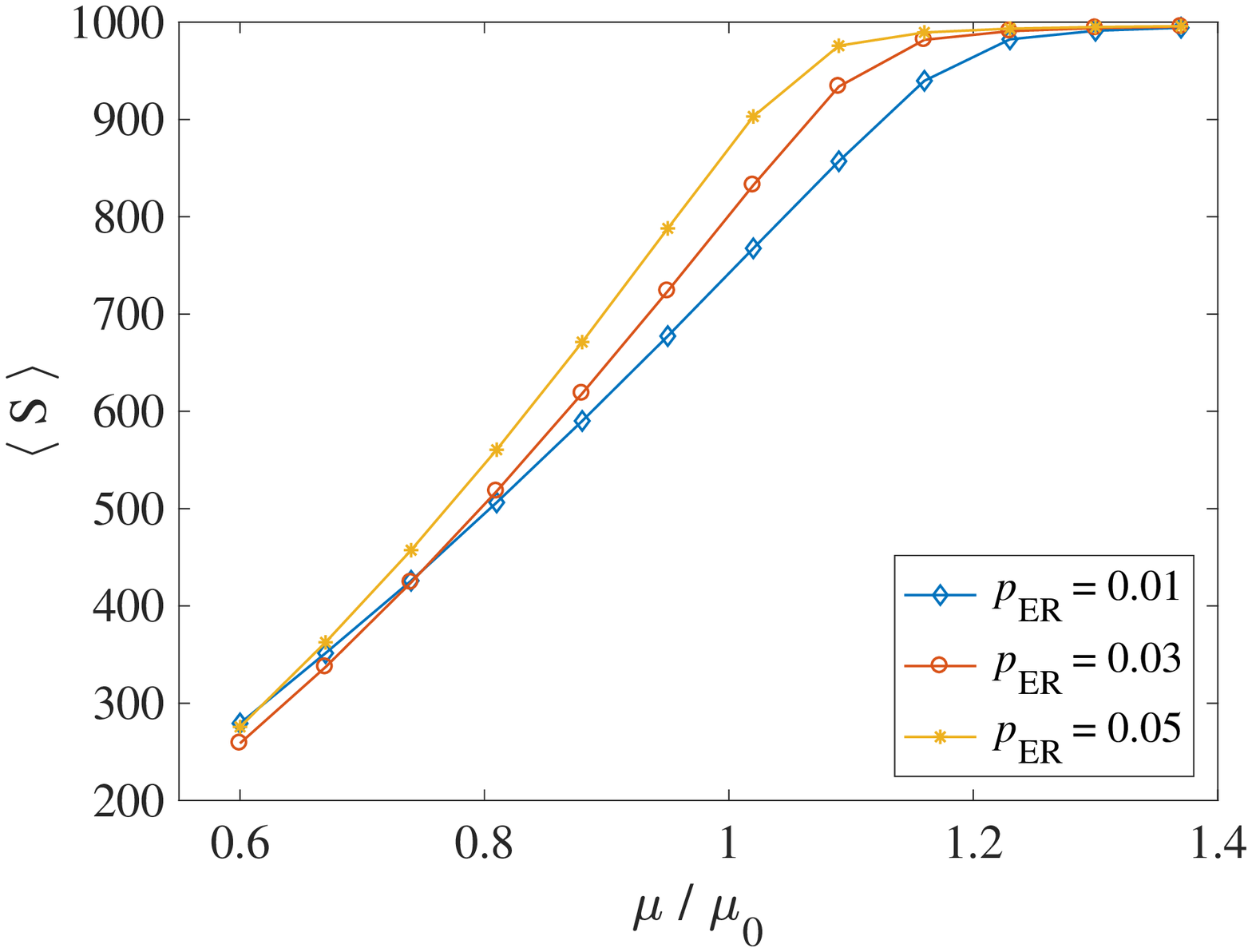}
\includegraphics[width=0.45\linewidth]{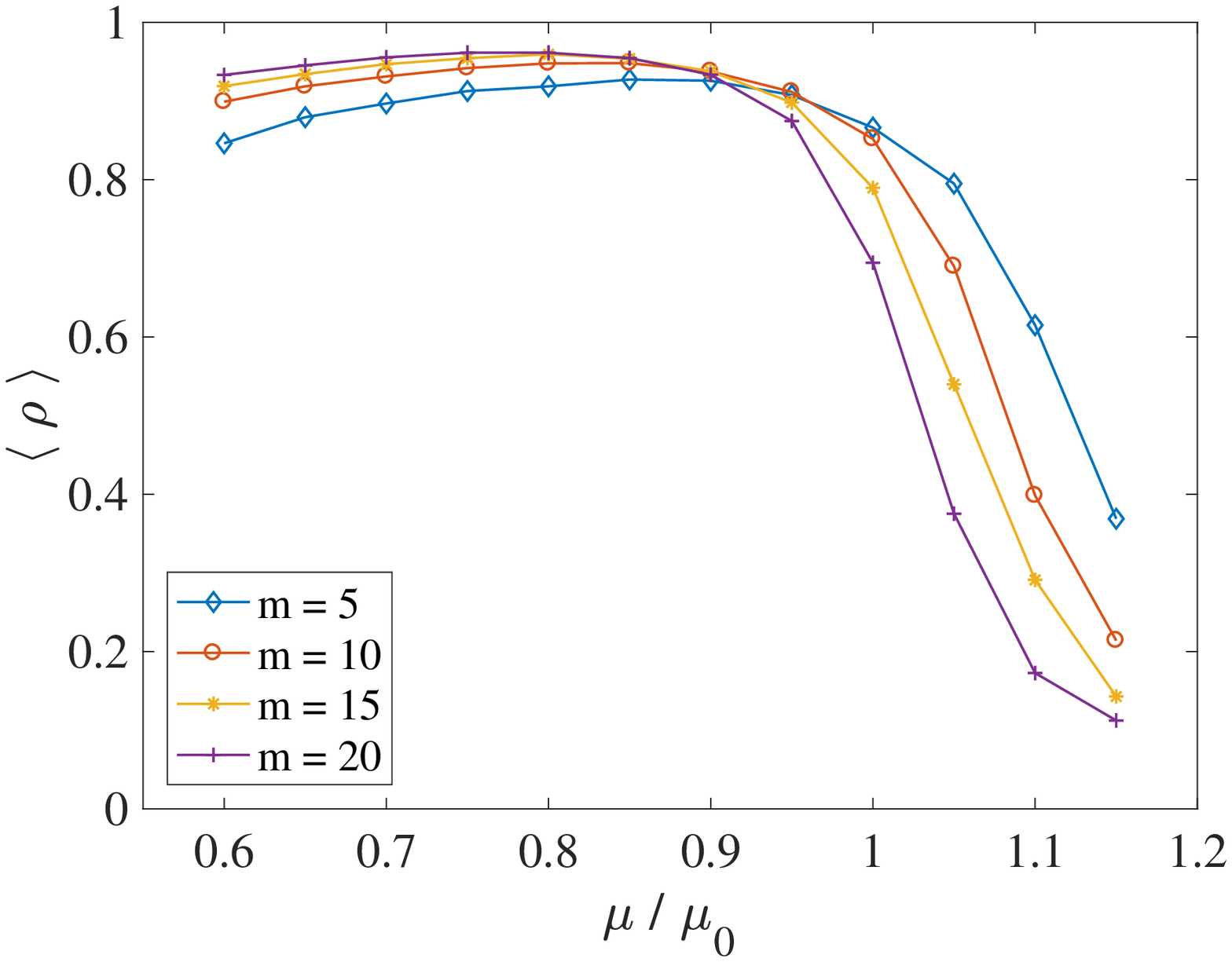}
\includegraphics[width=0.45\linewidth]{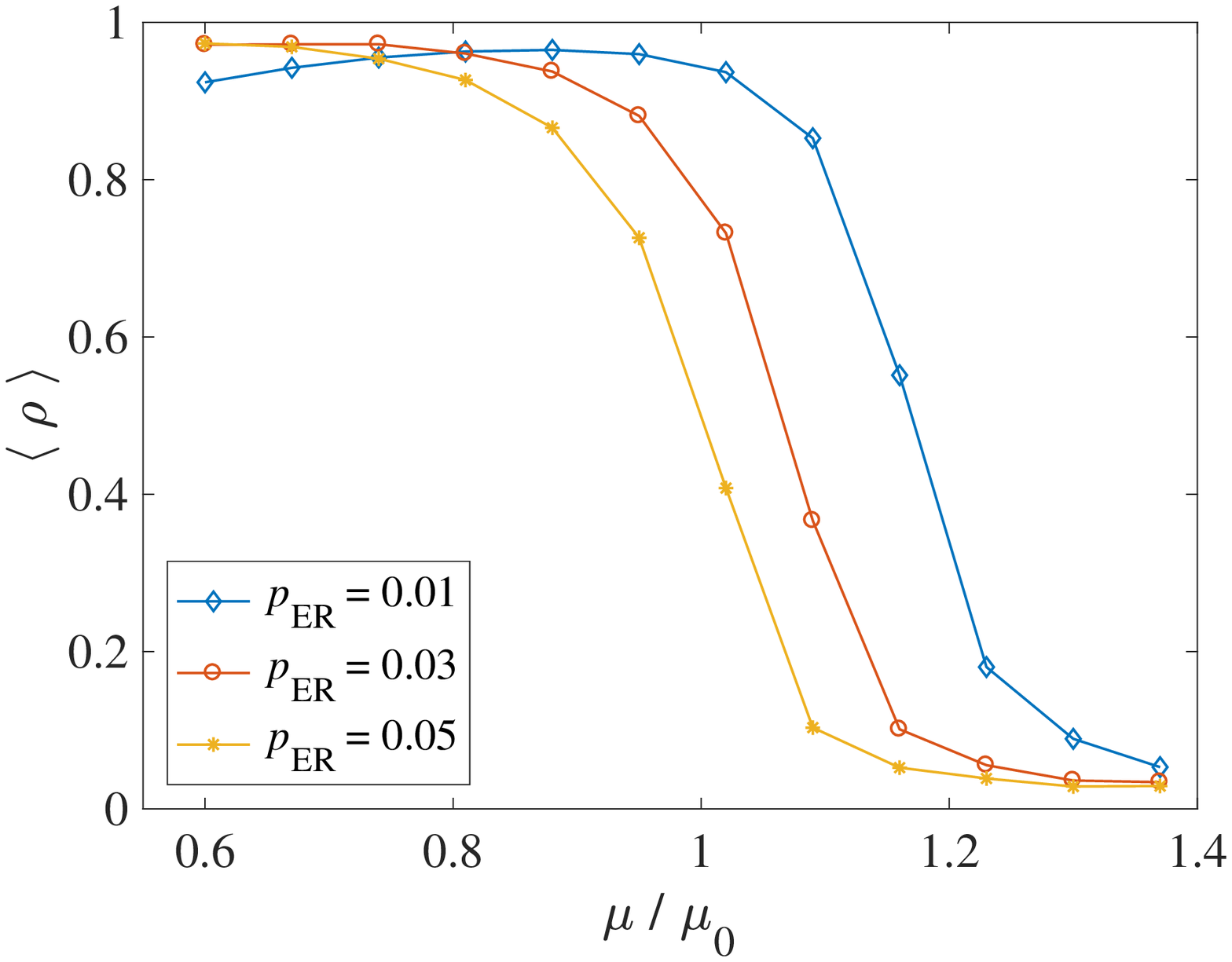}
\caption{\textbf{Discrete-time epidemic model on complex networks at the end of epidemic.} In all panels ten different complex networks with 1000 nodes are considered. The seed of generating the BA networks, $m$, and the link probability for ER networks $p_{\text{ER}}$ is given in the inset. In the top panels are shown the average number of susceptible individuals $\langle S \rangle $, while at bottom are average correlation coefficients $\langle \rho \rangle $ between the number of recovered individuals and the principal eigenvector of the respective adjacency matrix. The horizontal axis is given in units of the critical value of the parameter $\mu_0$ at the epidemic threshold which is calculated for each network separately.
\label{fig:BA_ER_Corr_Susc}}
\end{figure*}

Next, we have considered disease spreading on Erd\H{o}s-R\'enyi (ER) \cite{erdos1960evolution} and Barab\'{a}si-Albert (BA) \cite{barabasi1999emergence} models of complex networks. Within the ER model, we have considered probability of existence of link between each pair of nodes $p_{\text{ER}}\in\{0.01, 0.03, 0.05\}$ and generated ten different networks for each case. For the BA complex networks we have taken four different values of the seed $m \in \{5, 10, 15, 20\}$. As for the random regular graphs, for each ER and BA network ten different initial conditions were considered. The  parameter values for $\alpha$, $\beta$, $\gamma$ and $\sigma$ were taken identical as for the random regular graphs. In the figure \ref{fig:BA_ER_Corr_Susc} are shown the average number of susceptible individuals and the correlation coefficient between the recovered probability vector and the principal eigenvector of the adjacency matrix at the end of epidemic. Here, the threshold value $\mu_0$ was calculated for each network separately from the equation
\begin{equation}
    \Lambda_{\max}\frac{\alpha\mu_0 + \beta\sigma}{\mu_0(\mu_0 + \sigma)} = 1, \label{eq:mu_quadratic_eq_comp_net}
\end{equation}
where $\Lambda_{\max}$ is the leading eigenvalue of the given network. 

In the figure \ref{fig:BA_ER_Corr_Susc} can be seen that as the parameter $\mu$ increases towards the critical value $\mu_0$, the number of susceptible at the endemic equilibrium approaches the total number of individuals, as it is expected. We remark that it is not equal to the total population even when the conditions of epidemic are not satisfied, because there is certain probability that the patient zero will infect some neighboring nodes. However, this is finite size effect, and in infinitely large network the fraction of infected individuals is expected to be infinitesimally small in general. We remind that, in the related case, when the epidemic threshold is barely passed, for networks in which the principal eigenvector of the adjacency matrix is localized, only infinitesimal fraction of the population will be affected \cite{goltsev2012localization}, although for general networks it will be finite. One can also notice that the results about the ER network look that it is more prone to epidemic. This deception appears because the horizontal axis is in the units of the threshold value $\mu_0$ and not the absolute terms. We emphasize that, as it is well known, for infinite size BA networks the respective leading eigenvalue is infinite, and thus the threshold value of the contagiousness parameter is vanishing \cite{pastor2001PRL}. In our analysis of such networks, where the parameter $\mu$ is chosen to be varied, its critical value $\mu_0$ diverges for infinite networks.

The rather high value of the correlation coefficient $\rho$, when disease is spreading suggests that indeed the principal eigenvector predicts the pattern of infection. When the epidemic is not possible, $\mu > \mu_0$, the correlation does not drop sharply, due to the finite size effects. Near the epidemic threshold there is nonzero probability of infecting the neighborhood by the initially exposed node, and particularly those with higher eigenvector centrality.

\begin{figure*}[t]
\includegraphics[width=0.45\linewidth]{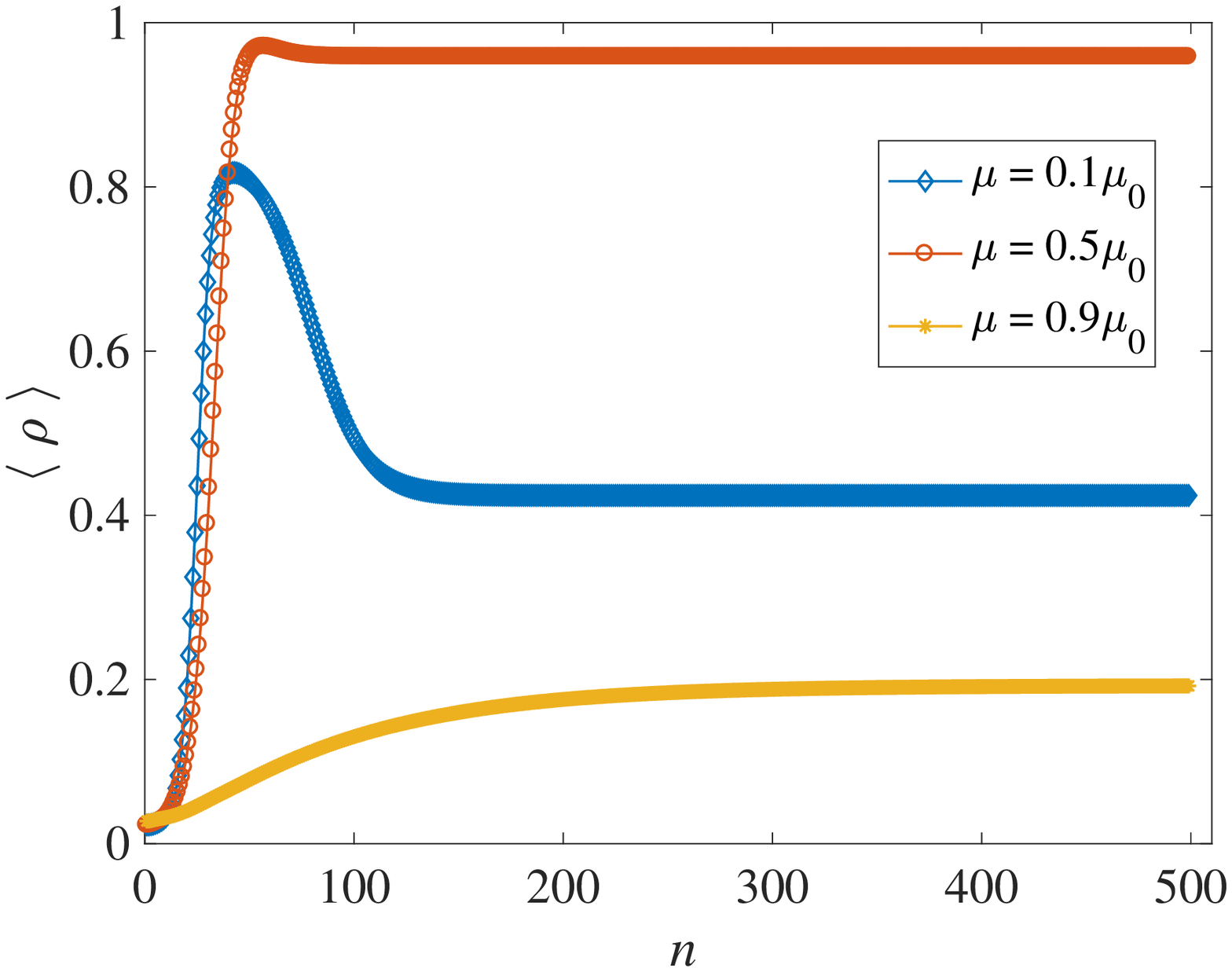}
\includegraphics[width=0.45\linewidth]{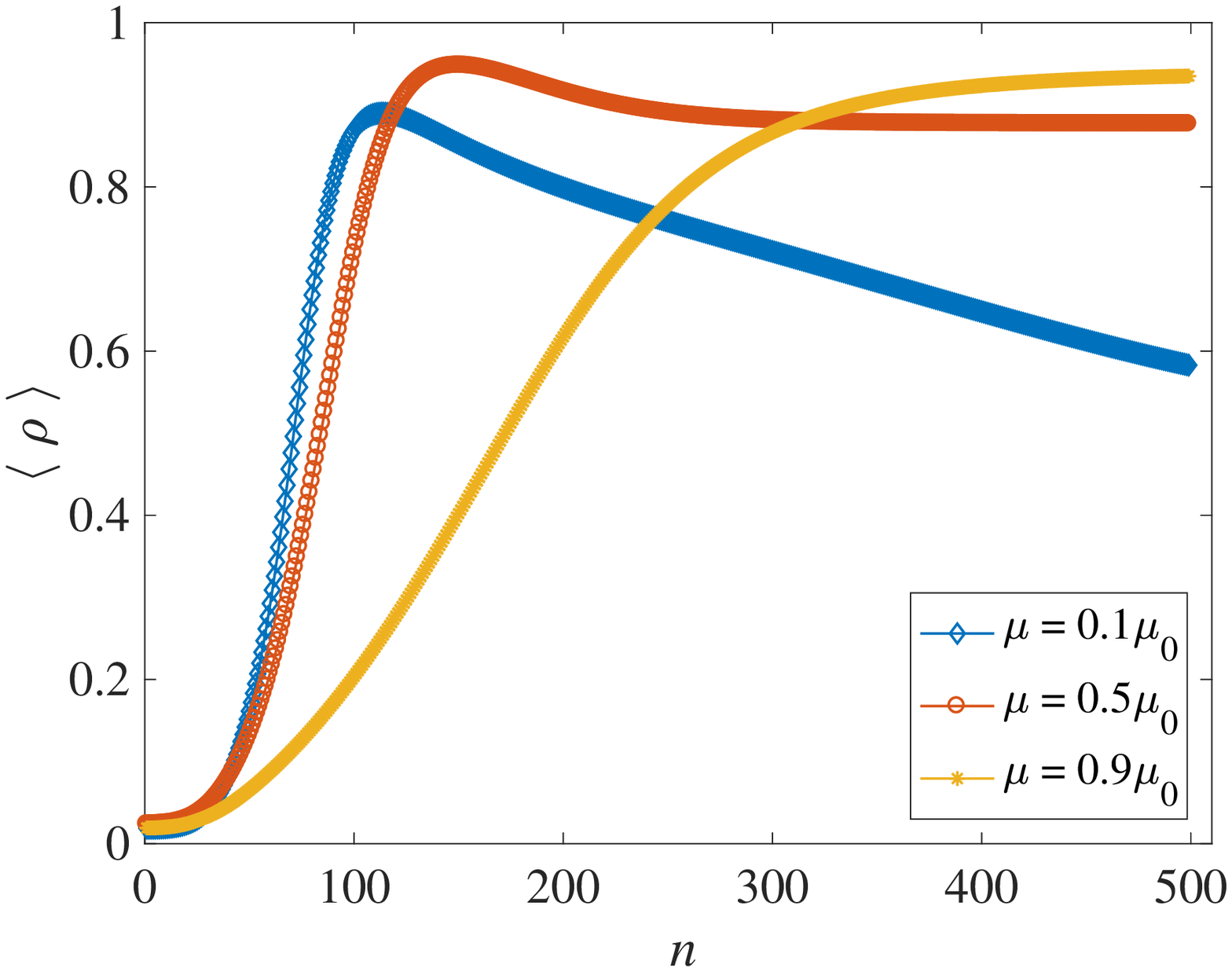}
\caption{\textbf{Evolution of correlation coefficient between the principal eigenvector of the adjacency matrix and the vector of probability of recovered state in ER (left panel) and BA (right panel) complex networks}. The considered networks have 1000 nodes. Each curve is obtained by averaging ten networks with the same parameters and ten randomly chosen initially infected nodes for each network.
\label{fig:Early_phase}}
\end{figure*}

We have finally studied the behaviour of the epidemics at the onset in order to verify which nodes bear the highest risk of contracting the disease. As common wisdom suggests, highly connected nodes, and particularly those with well connected neighbors are most risky ones -- just as the eigenvector centrality ranks the nodes. For that reason we have calculated the evolution of the correlation coefficient between the principal eigenvector of the adjacency matrix and the probability vector of the recovered state as the epidemic unfolds. In the figure \ref{fig:Early_phase} is shown the correlation coefficient as function of time. For the BA network shown at right the parameters have the same values as previously, while for the ER network (at left panel) $\alpha = 0.05$ and $\beta =0.04$, while $\gamma$ and $\sigma$ are the same as in the other simulations. One can note that generally in the early stages of the disease outbreak very high correlation is achieved, which confirms that the eigenvector centrality predicts rather well the riskiness of contraction of the pathogen. As the epidemics fades out the correlation might drop, because for certain parameter combinations majority of population has high chance of becoming infected and this infection pattern can differ significantly from the predictions by the principal eigenvector of the adjacency matrix. However, the lowest curve for the ER network model shows that this is not always happening. In such situation, when epidemic is barely possible, only small fraction of population can be affected, particularly those which are close to the patient zero. This observation suggests further investigation of the pattern of risk in case of such small outbreaks.

\section{Conclusions}

We have studied SEAIR epidemic spreading model aimed to capture the contagiousness features of COVID-19. Theoretical analysis were made for the compartmental version as well as for discrete-time epidemic spreading on random regular graphs and complex networks. For the compartmental model the epidemic threshold was found and it was shown that it also determines the emergence of endemic equilibrium as well. When the contagiousness is weak, we have shown that for random regular graphs and complex networks the epidemic threshold obtained from stability analysis of the disease-free state depends in a similar way on the model parameters. As is known for many other disease spreading models, the epidemic threshold was obtained to depend on the leading eigenvalue of the adjacency matrix. We have also demonstrated that when endemic equilibrium exists, it is linearly stable in the three considered models. The theoretical analysis in this work has shown that the risk for infection of certain node is dependent on its eigenvector centrality. In early stages of epidemics, the eigenvector centrality points which nodes are most likely to be first to contract the disease, while in case of mild epidemic on complex network, it shows which nodes have more risk to contract the disease during the whole course of the epidemic. 

The analysis of the linear stability was based on two approaches. By appropriately organizing the probabilities of various states as dynamical variables it was obtained the Jacobian matrix of the equilibria which has form that allows analytical treatment. The first approach was based on applying Schur's determinant identity which lead to result that the nontrivial eigenvalues of the Jacobian are related to those of the adjacency matrix. In the second approach we have furthermore shown that eigenvectors corresponding to the nontrivial eigenvalues of the Jacobian are combinations of scaled eigenvectors of the adjacency matrix. We believe that these two techniques can be applied in a range of studies where multidimensional dynamical systems interact through complex topology of contacts.

Although the motivation for studying the SEAIR epidemic spreading model was the COVID-19 pandemic, we did not make any testing about its relevance on real data. Naturally, it is the first study which should follow this one. One of the key issues would be inference of the fraction of the population which has contracted the disease, but has not shown symptoms at all. This could help in estimating the likelihood of reappearance of the epidemic and its possible size, once it weakens. Since, in general, contagiousness parameters change during epidemic, testing the validity of the relationships for the fraction of susceptible individuals at the end of epidemic might not be easy task. However, at the early phase of an epidemic these parameters could be considered as constant. Then, by using real data, it could be verified how well the eigenvector centrality anticipates which individuals bear highest risk of infection. If it proves to be useful predictor, then a follow up is investigation of its relevance to planning of vaccination and other protective measures.

\begin{acknowledgments}
This research was partially supported by the Faculty of Computer Science and Engineering, at the SS Cyril and Methodius University in Skopje, Macedonia. 
\end{acknowledgments}

\bibliography{DTEpidemic}

\begin{thebibliography}{46}%
\makeatletter
\providecommand \@ifxundefined [1]{%
 \@ifx{#1\undefined}
}%
\providecommand \@ifnum [1]{%
 \ifnum #1\expandafter \@firstoftwo
 \else \expandafter \@secondoftwo
 \fi
}%
\providecommand \@ifx [1]{%
 \ifx #1\expandafter \@firstoftwo
 \else \expandafter \@secondoftwo
 \fi
}%
\providecommand \natexlab [1]{#1}%
\providecommand \enquote  [1]{``#1''}%
\providecommand \bibnamefont  [1]{#1}%
\providecommand \bibfnamefont [1]{#1}%
\providecommand \citenamefont [1]{#1}%
\providecommand \href@noop [0]{\@secondoftwo}%
\providecommand \href [0]{\begingroup \@sanitize@url \@href}%
\providecommand \@href[1]{\@@startlink{#1}\@@href}%
\providecommand \@@href[1]{\endgroup#1\@@endlink}%
\providecommand \@sanitize@url [0]{\catcode `\\12\catcode `\$12\catcode
  `\&12\catcode `\#12\catcode `\^12\catcode `\_12\catcode `\%12\relax}%
\providecommand \@@startlink[1]{}%
\providecommand \@@endlink[0]{}%
\providecommand \url  [0]{\begingroup\@sanitize@url \@url }%
\providecommand \@url [1]{\endgroup\@href {#1}{\urlprefix }}%
\providecommand \urlprefix  [0]{URL }%
\providecommand \Eprint [0]{\href }%
\providecommand \doibase [0]{http://dx.doi.org/}%
\providecommand \selectlanguage [0]{\@gobble}%
\providecommand \bibinfo  [0]{\@secondoftwo}%
\providecommand \bibfield  [0]{\@secondoftwo}%
\providecommand \translation [1]{[#1]}%
\providecommand \BibitemOpen [0]{}%
\providecommand \bibitemStop [0]{}%
\providecommand \bibitemNoStop [0]{.\EOS\space}%
\providecommand \EOS [0]{\spacefactor3000\relax}%
\providecommand \BibitemShut  [1]{\csname bibitem#1\endcsname}%
\let\auto@bib@innerbib\@empty
\bibitem [{\citenamefont {Bernoulli}(1760)}]{bernoulli1760essai}%
  \BibitemOpen
  \bibfield  {author} {\bibinfo {author} {\bibfnamefont {D.}~\bibnamefont
  {Bernoulli}},\ }\href@noop {} {\bibfield  {journal} {\bibinfo  {journal}
  {Histoire de l'Acad., Roy. Sci.(Paris) avec Mem}\ ,\ \bibinfo {pages} {1}}
  (\bibinfo {year} {1760})}\BibitemShut {NoStop}%
\bibitem [{\citenamefont {Hamer}(1906)}]{hamer1906epidemic}%
  \BibitemOpen
  \bibfield  {author} {\bibinfo {author} {\bibfnamefont {W.}~\bibnamefont
  {Hamer}},\ }\href@noop {} {\bibfield  {journal} {\bibinfo  {journal}
  {Lancet}\ }\textbf {\bibinfo {volume} {1}},\ \bibinfo {pages} {733} (\bibinfo
  {year} {1906})}\BibitemShut {NoStop}%
\bibitem [{\citenamefont {Ross}(1911)}]{ross1911prevention}%
  \BibitemOpen
  \bibfield  {author} {\bibinfo {author} {\bibfnamefont {R.}~\bibnamefont
  {Ross}},\ }\href@noop {} {\emph {\bibinfo {title} {The prevention of
  malaria}}}\ (\bibinfo  {publisher} {Murray},\ \bibinfo {year}
  {1911})\BibitemShut {NoStop}%
\bibitem [{\citenamefont {Kermack}\ and\ \citenamefont
  {McKendrick}(1927)}]{kermack1927contribution}%
  \BibitemOpen
  \bibfield  {author} {\bibinfo {author} {\bibfnamefont {W.~O.}\ \bibnamefont
  {Kermack}}\ and\ \bibinfo {author} {\bibfnamefont {A.~G.}\ \bibnamefont
  {McKendrick}},\ }\href@noop {} {\bibfield  {journal} {\bibinfo  {journal}
  {Proc. R. Soc. Lond. Series A, Mathematical and Physical Sciences}\ }\textbf
  {\bibinfo {volume} {115}},\ \bibinfo {pages} {700} (\bibinfo {year}
  {1927})}\BibitemShut {NoStop}%
\bibitem [{\citenamefont {Pastor-Satorras}\ and\ \citenamefont
  {Vespignani}(2001{\natexlab{a}})}]{pastor2001PRE}%
  \BibitemOpen
  \bibfield  {author} {\bibinfo {author} {\bibfnamefont {R.}~\bibnamefont
  {Pastor-Satorras}}\ and\ \bibinfo {author} {\bibfnamefont {A.}~\bibnamefont
  {Vespignani}},\ }\href@noop {} {\bibfield  {journal} {\bibinfo  {journal}
  {Phys. Rev. E}\ }\textbf {\bibinfo {volume} {63}},\ \bibinfo {pages} {066117}
  (\bibinfo {year} {2001}{\natexlab{a}})}\BibitemShut {NoStop}%
\bibitem [{\citenamefont {Pastor-Satorras}\ and\ \citenamefont
  {Vespignani}(2001{\natexlab{b}})}]{pastor2001PRL}%
  \BibitemOpen
  \bibfield  {author} {\bibinfo {author} {\bibfnamefont {R.}~\bibnamefont
  {Pastor-Satorras}}\ and\ \bibinfo {author} {\bibfnamefont {A.}~\bibnamefont
  {Vespignani}},\ }\href@noop {} {\bibfield  {journal} {\bibinfo  {journal}
  {Phys. Rev. Lett.}\ }\textbf {\bibinfo {volume} {86}},\ \bibinfo {pages}
  {3200} (\bibinfo {year} {2001}{\natexlab{b}})}\BibitemShut {NoStop}%
\bibitem [{\citenamefont {Newman}(2002)}]{newman2002spread}%
  \BibitemOpen
  \bibfield  {author} {\bibinfo {author} {\bibfnamefont {M.~E.}\ \bibnamefont
  {Newman}},\ }\href@noop {} {\bibfield  {journal} {\bibinfo  {journal} {Phys.
  Rev. E}\ }\textbf {\bibinfo {volume} {66}},\ \bibinfo {pages} {016128}
  (\bibinfo {year} {2002})}\BibitemShut {NoStop}%
\bibitem [{\citenamefont {Vespignani}(2012)}]{vespignani2012modelling}%
  \BibitemOpen
  \bibfield  {author} {\bibinfo {author} {\bibfnamefont {A.}~\bibnamefont
  {Vespignani}},\ }\href@noop {} {\bibfield  {journal} {\bibinfo  {journal}
  {Nat. Phys.}\ }\textbf {\bibinfo {volume} {8}},\ \bibinfo {pages} {32}
  (\bibinfo {year} {2012})}\BibitemShut {NoStop}%
\bibitem [{\citenamefont {Pastor-Satorras}\ \emph {et~al.}(2015)\citenamefont
  {Pastor-Satorras}, \citenamefont {Castellano}, \citenamefont {Van~Mieghem},\
  and\ \citenamefont {Vespignani}}]{pastor2015epidemic}%
  \BibitemOpen
  \bibfield  {author} {\bibinfo {author} {\bibfnamefont {R.}~\bibnamefont
  {Pastor-Satorras}}, \bibinfo {author} {\bibfnamefont {C.}~\bibnamefont
  {Castellano}}, \bibinfo {author} {\bibfnamefont {P.}~\bibnamefont
  {Van~Mieghem}}, \ and\ \bibinfo {author} {\bibfnamefont {A.}~\bibnamefont
  {Vespignani}},\ }\href@noop {} {\bibfield  {journal} {\bibinfo  {journal}
  {Rev. Mod. Phys.}\ }\textbf {\bibinfo {volume} {87}},\ \bibinfo {pages} {925}
  (\bibinfo {year} {2015})}\BibitemShut {NoStop}%
\bibitem [{\citenamefont {de~Arruda}\ \emph {et~al.}(2018)\citenamefont
  {de~Arruda}, \citenamefont {Rodrigues},\ and\ \citenamefont
  {Moreno}}]{de2018fundamentals}%
  \BibitemOpen
  \bibfield  {author} {\bibinfo {author} {\bibfnamefont {G.~F.}\ \bibnamefont
  {de~Arruda}}, \bibinfo {author} {\bibfnamefont {F.~A.}\ \bibnamefont
  {Rodrigues}}, \ and\ \bibinfo {author} {\bibfnamefont {Y.}~\bibnamefont
  {Moreno}},\ }\href@noop {} {\bibfield  {journal} {\bibinfo  {journal} {Phys.
  Rep.}\ }\textbf {\bibinfo {volume} {756}},\ \bibinfo {pages} {1} (\bibinfo
  {year} {2018})}\BibitemShut {NoStop}%
\bibitem [{\citenamefont {Watts}\ and\ \citenamefont
  {Strogatz}(1998)}]{watts1998collective}%
  \BibitemOpen
  \bibfield  {author} {\bibinfo {author} {\bibfnamefont {D.~J.}\ \bibnamefont
  {Watts}}\ and\ \bibinfo {author} {\bibfnamefont {S.~H.}\ \bibnamefont
  {Strogatz}},\ }\href@noop {} {\bibfield  {journal} {\bibinfo  {journal}
  {Nature}\ }\textbf {\bibinfo {volume} {393}},\ \bibinfo {pages} {440}
  (\bibinfo {year} {1998})}\BibitemShut {NoStop}%
\bibitem [{\citenamefont {Barab{\'a}si}\ and\ \citenamefont
  {Albert}(1999)}]{barabasi1999emergence}%
  \BibitemOpen
  \bibfield  {author} {\bibinfo {author} {\bibfnamefont {A.-L.}\ \bibnamefont
  {Barab{\'a}si}}\ and\ \bibinfo {author} {\bibfnamefont {R.}~\bibnamefont
  {Albert}},\ }\href@noop {} {\bibfield  {journal} {\bibinfo  {journal}
  {Science}\ }\textbf {\bibinfo {volume} {286}},\ \bibinfo {pages} {509}
  (\bibinfo {year} {1999})}\BibitemShut {NoStop}%
\bibitem [{\citenamefont {Bogun{\'a}}\ \emph {et~al.}(2003)\citenamefont
  {Bogun{\'a}}, \citenamefont {Pastor-Satorras},\ and\ \citenamefont
  {Vespignani}}]{boguna2003absence}%
  \BibitemOpen
  \bibfield  {author} {\bibinfo {author} {\bibfnamefont {M.}~\bibnamefont
  {Bogun{\'a}}}, \bibinfo {author} {\bibfnamefont {R.}~\bibnamefont
  {Pastor-Satorras}}, \ and\ \bibinfo {author} {\bibfnamefont {A.}~\bibnamefont
  {Vespignani}},\ }\href@noop {} {\bibfield  {journal} {\bibinfo  {journal}
  {Phys. Rev. Lett.}\ }\textbf {\bibinfo {volume} {90}},\ \bibinfo {pages}
  {028701} (\bibinfo {year} {2003})}\BibitemShut {NoStop}%
\bibitem [{\citenamefont {Wang}\ \emph {et~al.}(2003)\citenamefont {Wang},
  \citenamefont {Chakrabarti}, \citenamefont {Wang},\ and\ \citenamefont
  {Faloutsos}}]{wang2003epidemic}%
  \BibitemOpen
  \bibfield  {author} {\bibinfo {author} {\bibfnamefont {Y.}~\bibnamefont
  {Wang}}, \bibinfo {author} {\bibfnamefont {D.}~\bibnamefont {Chakrabarti}},
  \bibinfo {author} {\bibfnamefont {C.}~\bibnamefont {Wang}}, \ and\ \bibinfo
  {author} {\bibfnamefont {C.}~\bibnamefont {Faloutsos}},\ }in\ \href@noop {}
  {\emph {\bibinfo {booktitle} {22nd International Symposium on Reliable
  Distributed Systems, 2003. Proceedings.}}}\ (\bibinfo {organization} {IEEE},\
  \bibinfo {year} {2003})\ pp.\ \bibinfo {pages} {25--34}\BibitemShut {NoStop}%
\bibitem [{\citenamefont {Van~Mieghem}\ \emph {et~al.}(2008)\citenamefont
  {Van~Mieghem}, \citenamefont {Omic},\ and\ \citenamefont
  {Kooij}}]{van2008virus}%
  \BibitemOpen
  \bibfield  {author} {\bibinfo {author} {\bibfnamefont {P.}~\bibnamefont
  {Van~Mieghem}}, \bibinfo {author} {\bibfnamefont {J.}~\bibnamefont {Omic}}, \
  and\ \bibinfo {author} {\bibfnamefont {R.}~\bibnamefont {Kooij}},\
  }\href@noop {} {\bibfield  {journal} {\bibinfo  {journal} {IEEE/ACM
  Transactions On Networking}\ }\textbf {\bibinfo {volume} {17}},\ \bibinfo
  {pages} {1} (\bibinfo {year} {2008})}\BibitemShut {NoStop}%
\bibitem [{\citenamefont {Goltsev}\ \emph {et~al.}(2012)\citenamefont
  {Goltsev}, \citenamefont {Dorogovtsev}, \citenamefont {Oliveira},\ and\
  \citenamefont {Mendes}}]{goltsev2012localization}%
  \BibitemOpen
  \bibfield  {author} {\bibinfo {author} {\bibfnamefont {A.~V.}\ \bibnamefont
  {Goltsev}}, \bibinfo {author} {\bibfnamefont {S.~N.}\ \bibnamefont
  {Dorogovtsev}}, \bibinfo {author} {\bibfnamefont {J.~G.}\ \bibnamefont
  {Oliveira}}, \ and\ \bibinfo {author} {\bibfnamefont {J.~F.}\ \bibnamefont
  {Mendes}},\ }\href@noop {} {\bibfield  {journal} {\bibinfo  {journal} {Phys.
  Rev. Lett.}\ }\textbf {\bibinfo {volume} {109}},\ \bibinfo {pages} {128702}
  (\bibinfo {year} {2012})}\BibitemShut {NoStop}%
\bibitem [{\citenamefont {de~Arruda}\ \emph {et~al.}(2017)\citenamefont
  {de~Arruda}, \citenamefont {Cozzo}, \citenamefont {Peixoto}, \citenamefont
  {Rodrigues},\ and\ \citenamefont {Moreno}}]{de2017disease}%
  \BibitemOpen
  \bibfield  {author} {\bibinfo {author} {\bibfnamefont {G.~F.}\ \bibnamefont
  {de~Arruda}}, \bibinfo {author} {\bibfnamefont {E.}~\bibnamefont {Cozzo}},
  \bibinfo {author} {\bibfnamefont {T.~P.}\ \bibnamefont {Peixoto}}, \bibinfo
  {author} {\bibfnamefont {F.~A.}\ \bibnamefont {Rodrigues}}, \ and\ \bibinfo
  {author} {\bibfnamefont {Y.}~\bibnamefont {Moreno}},\ }\href@noop {}
  {\bibfield  {journal} {\bibinfo  {journal} {Phys. Rev. X}\ }\textbf {\bibinfo
  {volume} {7}},\ \bibinfo {pages} {011014} (\bibinfo {year}
  {2017})}\BibitemShut {NoStop}%
\bibitem [{\citenamefont {Wang}\ \emph {et~al.}(2020)\citenamefont {Wang},
  \citenamefont {Xia}, \citenamefont {Chen},\ and\ \citenamefont
  {Chen}}]{wang2020epidemic}%
  \BibitemOpen
  \bibfield  {author} {\bibinfo {author} {\bibfnamefont {Z.}~\bibnamefont
  {Wang}}, \bibinfo {author} {\bibfnamefont {C.}~\bibnamefont {Xia}}, \bibinfo
  {author} {\bibfnamefont {Z.}~\bibnamefont {Chen}}, \ and\ \bibinfo {author}
  {\bibfnamefont {G.}~\bibnamefont {Chen}},\ }\href@noop {} {\bibfield
  {journal} {\bibinfo  {journal} {IEEE Trans. Cybern.}\ } (\bibinfo {year}
  {2020})}\BibitemShut {NoStop}%
\bibitem [{\citenamefont {Van~Mieghem}\ and\ \citenamefont {Van~de
  Bovenkamp}(2013)}]{van2013non}%
  \BibitemOpen
  \bibfield  {author} {\bibinfo {author} {\bibfnamefont {P.}~\bibnamefont
  {Van~Mieghem}}\ and\ \bibinfo {author} {\bibfnamefont {R.}~\bibnamefont
  {Van~de Bovenkamp}},\ }\href@noop {} {\bibfield  {journal} {\bibinfo
  {journal} {Phys. Rev. Lett.}\ }\textbf {\bibinfo {volume} {110}},\ \bibinfo
  {pages} {108701} (\bibinfo {year} {2013})}\BibitemShut {NoStop}%
\bibitem [{\citenamefont {Starnini}\ \emph {et~al.}(2017)\citenamefont
  {Starnini}, \citenamefont {Gleeson},\ and\ \citenamefont
  {Bogu{\~n}{\'a}}}]{starnini2017equivalence}%
  \BibitemOpen
  \bibfield  {author} {\bibinfo {author} {\bibfnamefont {M.}~\bibnamefont
  {Starnini}}, \bibinfo {author} {\bibfnamefont {J.~P.}\ \bibnamefont
  {Gleeson}}, \ and\ \bibinfo {author} {\bibfnamefont {M.}~\bibnamefont
  {Bogu{\~n}{\'a}}},\ }\href@noop {} {\bibfield  {journal} {\bibinfo  {journal}
  {Phys. Rev. Lett.}\ }\textbf {\bibinfo {volume} {118}},\ \bibinfo {pages}
  {128301} (\bibinfo {year} {2017})}\BibitemShut {NoStop}%
\bibitem [{\citenamefont {Xia}\ \emph {et~al.}(2013)\citenamefont {Xia},
  \citenamefont {Wang}, \citenamefont {Sanz}, \citenamefont {Meloni},\ and\
  \citenamefont {Moreno}}]{xia2013effects}%
  \BibitemOpen
  \bibfield  {author} {\bibinfo {author} {\bibfnamefont {C.-y.}\ \bibnamefont
  {Xia}}, \bibinfo {author} {\bibfnamefont {Z.}~\bibnamefont {Wang}}, \bibinfo
  {author} {\bibfnamefont {J.}~\bibnamefont {Sanz}}, \bibinfo {author}
  {\bibfnamefont {S.}~\bibnamefont {Meloni}}, \ and\ \bibinfo {author}
  {\bibfnamefont {Y.}~\bibnamefont {Moreno}},\ }\href@noop {} {\bibfield
  {journal} {\bibinfo  {journal} {Physica A}\ }\textbf {\bibinfo {volume}
  {392}},\ \bibinfo {pages} {1577} (\bibinfo {year} {2013})}\BibitemShut
  {NoStop}%
\bibitem [{\citenamefont {Xia}\ \emph {et~al.}(2012)\citenamefont {Xia},
  \citenamefont {Wang}, \citenamefont {Sun},\ and\ \citenamefont
  {Wang}}]{xia2012sir}%
  \BibitemOpen
  \bibfield  {author} {\bibinfo {author} {\bibfnamefont {C.}~\bibnamefont
  {Xia}}, \bibinfo {author} {\bibfnamefont {L.}~\bibnamefont {Wang}}, \bibinfo
  {author} {\bibfnamefont {S.}~\bibnamefont {Sun}}, \ and\ \bibinfo {author}
  {\bibfnamefont {J.}~\bibnamefont {Wang}},\ }\href@noop {} {\bibfield
  {journal} {\bibinfo  {journal} {Nonlinear Dyn.}\ }\textbf {\bibinfo {volume}
  {69}},\ \bibinfo {pages} {927} (\bibinfo {year} {2012})}\BibitemShut
  {NoStop}%
\bibitem [{\citenamefont {Hethcote}(2000)}]{hethcote2000mathematics}%
  \BibitemOpen
  \bibfield  {author} {\bibinfo {author} {\bibfnamefont {H.~W.}\ \bibnamefont
  {Hethcote}},\ }\href@noop {} {\bibfield  {journal} {\bibinfo  {journal} {SIAM
  Rev.}\ }\textbf {\bibinfo {volume} {42}},\ \bibinfo {pages} {599} (\bibinfo
  {year} {2000})}\BibitemShut {NoStop}%
\bibitem [{\citenamefont {Prakash}\ \emph {et~al.}(2010)\citenamefont
  {Prakash}, \citenamefont {Chakrabarti}, \citenamefont {Faloutsos},
  \citenamefont {Valler},\ and\ \citenamefont {Faloutsos}}]{prakash2010got}%
  \BibitemOpen
  \bibfield  {author} {\bibinfo {author} {\bibfnamefont {B.~A.}\ \bibnamefont
  {Prakash}}, \bibinfo {author} {\bibfnamefont {D.}~\bibnamefont
  {Chakrabarti}}, \bibinfo {author} {\bibfnamefont {M.}~\bibnamefont
  {Faloutsos}}, \bibinfo {author} {\bibfnamefont {N.}~\bibnamefont {Valler}}, \
  and\ \bibinfo {author} {\bibfnamefont {C.}~\bibnamefont {Faloutsos}},\
  }\href@noop {} {\bibfield  {journal} {\bibinfo  {journal} {arXiv preprint
  arXiv:1004.0060}\ } (\bibinfo {year} {2010})}\BibitemShut {NoStop}%
\bibitem [{\citenamefont {He}\ \emph {et~al.}(2020)\citenamefont {He},
  \citenamefont {Lau}, \citenamefont {Wu}, \citenamefont {Deng}, \citenamefont
  {Wang}, \citenamefont {Hao}, \citenamefont {Lau}, \citenamefont {Wong},
  \citenamefont {Guan}, \citenamefont {Tan} \emph {et~al.}}]{he2020temporal}%
  \BibitemOpen
  \bibfield  {author} {\bibinfo {author} {\bibfnamefont {X.}~\bibnamefont
  {He}}, \bibinfo {author} {\bibfnamefont {E.~H.}\ \bibnamefont {Lau}},
  \bibinfo {author} {\bibfnamefont {P.}~\bibnamefont {Wu}}, \bibinfo {author}
  {\bibfnamefont {X.}~\bibnamefont {Deng}}, \bibinfo {author} {\bibfnamefont
  {J.}~\bibnamefont {Wang}}, \bibinfo {author} {\bibfnamefont {X.}~\bibnamefont
  {Hao}}, \bibinfo {author} {\bibfnamefont {Y.~C.}\ \bibnamefont {Lau}},
  \bibinfo {author} {\bibfnamefont {J.~Y.}\ \bibnamefont {Wong}}, \bibinfo
  {author} {\bibfnamefont {Y.}~\bibnamefont {Guan}}, \bibinfo {author}
  {\bibfnamefont {X.}~\bibnamefont {Tan}},  \emph {et~al.},\ }\href@noop {}
  {\bibfield  {journal} {\bibinfo  {journal} {Nat. Med.}\ }\textbf {\bibinfo
  {volume} {26}},\ \bibinfo {pages} {672} (\bibinfo {year} {2020})}\BibitemShut
  {NoStop}%
\bibitem [{\citenamefont {Du}\ \emph {et~al.}(2020)\citenamefont {Du},
  \citenamefont {Xu}, \citenamefont {Wu}, \citenamefont {Wang}, \citenamefont
  {Cowling},\ and\ \citenamefont {Meyers}}]{du2020serial}%
  \BibitemOpen
  \bibfield  {author} {\bibinfo {author} {\bibfnamefont {Z.}~\bibnamefont
  {Du}}, \bibinfo {author} {\bibfnamefont {X.}~\bibnamefont {Xu}}, \bibinfo
  {author} {\bibfnamefont {Y.}~\bibnamefont {Wu}}, \bibinfo {author}
  {\bibfnamefont {L.}~\bibnamefont {Wang}}, \bibinfo {author} {\bibfnamefont
  {B.~J.}\ \bibnamefont {Cowling}}, \ and\ \bibinfo {author} {\bibfnamefont
  {L.~A.}\ \bibnamefont {Meyers}},\ }\href@noop {} {\bibfield  {journal}
  {\bibinfo  {journal} {Emerg. Infect. Dis.}\ }\textbf {\bibinfo {volume}
  {26}},\ \bibinfo {pages} {1341} (\bibinfo {year} {2020})}\BibitemShut
  {NoStop}%
\bibitem [{\citenamefont {Lauer}\ \emph {et~al.}(2020)\citenamefont {Lauer},
  \citenamefont {Grantz}, \citenamefont {Bi}, \citenamefont {Jones},
  \citenamefont {Zheng}, \citenamefont {Meredith}, \citenamefont {Azman},
  \citenamefont {Reich},\ and\ \citenamefont {Lessler}}]{lauer2020incubation}%
  \BibitemOpen
  \bibfield  {author} {\bibinfo {author} {\bibfnamefont {S.~A.}\ \bibnamefont
  {Lauer}}, \bibinfo {author} {\bibfnamefont {K.~H.}\ \bibnamefont {Grantz}},
  \bibinfo {author} {\bibfnamefont {Q.}~\bibnamefont {Bi}}, \bibinfo {author}
  {\bibfnamefont {F.~K.}\ \bibnamefont {Jones}}, \bibinfo {author}
  {\bibfnamefont {Q.}~\bibnamefont {Zheng}}, \bibinfo {author} {\bibfnamefont
  {H.~R.}\ \bibnamefont {Meredith}}, \bibinfo {author} {\bibfnamefont {A.~S.}\
  \bibnamefont {Azman}}, \bibinfo {author} {\bibfnamefont {N.~G.}\ \bibnamefont
  {Reich}}, \ and\ \bibinfo {author} {\bibfnamefont {J.}~\bibnamefont
  {Lessler}},\ }\href@noop {} {\bibfield  {journal} {\bibinfo  {journal} {Ann.
  Intern. Med.}\ }\textbf {\bibinfo {volume} {172}},\ \bibinfo {pages} {577}
  (\bibinfo {year} {2020})}\BibitemShut {NoStop}%
\bibitem [{\citenamefont {Di~Domenico}\ \emph {et~al.}(2020)\citenamefont
  {Di~Domenico}, \citenamefont {Pullano}, \citenamefont {Sabbatini},
  \citenamefont {Bo{\"e}lle},\ and\ \citenamefont {Colizza}}]{di2020expected}%
  \BibitemOpen
  \bibfield  {author} {\bibinfo {author} {\bibfnamefont {L.}~\bibnamefont
  {Di~Domenico}}, \bibinfo {author} {\bibfnamefont {G.}~\bibnamefont
  {Pullano}}, \bibinfo {author} {\bibfnamefont {C.~E.}\ \bibnamefont
  {Sabbatini}}, \bibinfo {author} {\bibfnamefont {P.-Y.}\ \bibnamefont
  {Bo{\"e}lle}}, \ and\ \bibinfo {author} {\bibfnamefont {V.}~\bibnamefont
  {Colizza}},\ }\href@noop {} {\bibfield  {journal} {\bibinfo  {journal}
  {medRxiv}\ } (\bibinfo {year} {2020})}\BibitemShut {NoStop}%
\bibitem [{\citenamefont {Aleta}\ \emph {et~al.}(2020)\citenamefont {Aleta},
  \citenamefont {Martin-Corral}, \citenamefont {y~Piontti}, \citenamefont
  {Ajelli}, \citenamefont {Litvinova}, \citenamefont {Chinazzi}, \citenamefont
  {Dean}, \citenamefont {Halloran}, \citenamefont {Longini~Jr}, \citenamefont
  {Merler} \emph {et~al.}}]{aleta2020modeling}%
  \BibitemOpen
  \bibfield  {author} {\bibinfo {author} {\bibfnamefont {A.}~\bibnamefont
  {Aleta}}, \bibinfo {author} {\bibfnamefont {D.}~\bibnamefont
  {Martin-Corral}}, \bibinfo {author} {\bibfnamefont {A.~P.}\ \bibnamefont
  {y~Piontti}}, \bibinfo {author} {\bibfnamefont {M.}~\bibnamefont {Ajelli}},
  \bibinfo {author} {\bibfnamefont {M.}~\bibnamefont {Litvinova}}, \bibinfo
  {author} {\bibfnamefont {M.}~\bibnamefont {Chinazzi}}, \bibinfo {author}
  {\bibfnamefont {N.~E.}\ \bibnamefont {Dean}}, \bibinfo {author}
  {\bibfnamefont {M.~E.}\ \bibnamefont {Halloran}}, \bibinfo {author}
  {\bibfnamefont {I.~M.}\ \bibnamefont {Longini~Jr}}, \bibinfo {author}
  {\bibfnamefont {S.}~\bibnamefont {Merler}},  \emph {et~al.},\ }\href@noop {}
  {\bibfield  {journal} {\bibinfo  {journal} {medRxiv}\ } (\bibinfo {year}
  {2020})}\BibitemShut {NoStop}%
\bibitem [{\citenamefont {Ndairou}\ \emph {et~al.}(2020)\citenamefont
  {Ndairou}, \citenamefont {Area}, \citenamefont {Nieto},\ and\ \citenamefont
  {Torres}}]{ndairou2020mathematical}%
  \BibitemOpen
  \bibfield  {author} {\bibinfo {author} {\bibfnamefont {F.}~\bibnamefont
  {Ndairou}}, \bibinfo {author} {\bibfnamefont {I.}~\bibnamefont {Area}},
  \bibinfo {author} {\bibfnamefont {J.~J.}\ \bibnamefont {Nieto}}, \ and\
  \bibinfo {author} {\bibfnamefont {D.~F.}\ \bibnamefont {Torres}},\
  }\href@noop {} {\bibfield  {journal} {\bibinfo  {journal} {Chaos Soliton.
  Fract.}\ }\textbf {\bibinfo {volume} {135}},\ \bibinfo {pages} {109846}
  (\bibinfo {year} {2020})}\BibitemShut {NoStop}%
\bibitem [{\citenamefont {Giordano}\ \emph {et~al.}(2020)\citenamefont
  {Giordano}, \citenamefont {Blanchini}, \citenamefont {Bruno}, \citenamefont
  {Colaneri}, \citenamefont {Di~Filippo}, \citenamefont {Di~Matteo},\ and\
  \citenamefont {Colaneri}}]{giordano2020modelling}%
  \BibitemOpen
  \bibfield  {author} {\bibinfo {author} {\bibfnamefont {G.}~\bibnamefont
  {Giordano}}, \bibinfo {author} {\bibfnamefont {F.}~\bibnamefont {Blanchini}},
  \bibinfo {author} {\bibfnamefont {R.}~\bibnamefont {Bruno}}, \bibinfo
  {author} {\bibfnamefont {P.}~\bibnamefont {Colaneri}}, \bibinfo {author}
  {\bibfnamefont {A.}~\bibnamefont {Di~Filippo}}, \bibinfo {author}
  {\bibfnamefont {A.}~\bibnamefont {Di~Matteo}}, \ and\ \bibinfo {author}
  {\bibfnamefont {M.}~\bibnamefont {Colaneri}},\ }\href@noop {} {\bibfield
  {journal} {\bibinfo  {journal} {Nat. Med.}\ ,\ \bibinfo {pages} {1}}
  (\bibinfo {year} {2020})}\BibitemShut {NoStop}%
\bibitem [{\citenamefont {Zhao}\ and\ \citenamefont
  {Chen}(2020)}]{zhao2020modeling}%
  \BibitemOpen
  \bibfield  {author} {\bibinfo {author} {\bibfnamefont {S.}~\bibnamefont
  {Zhao}}\ and\ \bibinfo {author} {\bibfnamefont {H.}~\bibnamefont {Chen}},\
  }\href@noop {} {\bibfield  {journal} {\bibinfo  {journal} {Quant. Biol.}\
  }\textbf {\bibinfo {volume} {8}},\ \bibinfo {pages} {11} (\bibinfo {year}
  {2020})}\BibitemShut {NoStop}%
\bibitem [{\citenamefont {Peng}\ \emph {et~al.}(2020)\citenamefont {Peng},
  \citenamefont {Yang}, \citenamefont {Zhang}, \citenamefont {Zhuge},\ and\
  \citenamefont {Hong}}]{peng2020epidemic}%
  \BibitemOpen
  \bibfield  {author} {\bibinfo {author} {\bibfnamefont {L.}~\bibnamefont
  {Peng}}, \bibinfo {author} {\bibfnamefont {W.}~\bibnamefont {Yang}}, \bibinfo
  {author} {\bibfnamefont {D.}~\bibnamefont {Zhang}}, \bibinfo {author}
  {\bibfnamefont {C.}~\bibnamefont {Zhuge}}, \ and\ \bibinfo {author}
  {\bibfnamefont {L.}~\bibnamefont {Hong}},\ }\href@noop {} {\bibfield
  {journal} {\bibinfo  {journal} {arXiv preprint arXiv:2002.06563}\ } (\bibinfo
  {year} {2020})}\BibitemShut {NoStop}%
\bibitem [{\citenamefont {Gatto}\ \emph {et~al.}(2020)\citenamefont {Gatto},
  \citenamefont {Bertuzzo}, \citenamefont {Mari}, \citenamefont {Miccoli},
  \citenamefont {Carraro}, \citenamefont {Casagrandi},\ and\ \citenamefont
  {Rinaldo}}]{gatto2020spread}%
  \BibitemOpen
  \bibfield  {author} {\bibinfo {author} {\bibfnamefont {M.}~\bibnamefont
  {Gatto}}, \bibinfo {author} {\bibfnamefont {E.}~\bibnamefont {Bertuzzo}},
  \bibinfo {author} {\bibfnamefont {L.}~\bibnamefont {Mari}}, \bibinfo {author}
  {\bibfnamefont {S.}~\bibnamefont {Miccoli}}, \bibinfo {author} {\bibfnamefont
  {L.}~\bibnamefont {Carraro}}, \bibinfo {author} {\bibfnamefont
  {R.}~\bibnamefont {Casagrandi}}, \ and\ \bibinfo {author} {\bibfnamefont
  {A.}~\bibnamefont {Rinaldo}},\ }\href@noop {} {\bibfield  {journal} {\bibinfo
   {journal} {PNAS}\ }\textbf {\bibinfo {volume} {117}},\ \bibinfo {pages}
  {10484} (\bibinfo {year} {2020})}\BibitemShut {NoStop}%
\bibitem [{\citenamefont {Nabi}(2020)}]{nabi2020forecasting}%
  \BibitemOpen
  \bibfield  {author} {\bibinfo {author} {\bibfnamefont {K.~N.}\ \bibnamefont
  {Nabi}},\ }\href@noop {} {\bibfield  {journal} {\bibinfo  {journal} {Chaos
  Soliton. Fract.}\ ,\ \bibinfo {pages} {110046}} (\bibinfo {year}
  {2020})}\BibitemShut {NoStop}%
\bibitem [{\citenamefont {Liu}\ \emph {et~al.}(2020)\citenamefont {Liu},
  \citenamefont {Magal}, \citenamefont {Seydi},\ and\ \citenamefont
  {Webb}}]{liu2020covid}%
  \BibitemOpen
  \bibfield  {author} {\bibinfo {author} {\bibfnamefont {Z.}~\bibnamefont
  {Liu}}, \bibinfo {author} {\bibfnamefont {P.}~\bibnamefont {Magal}}, \bibinfo
  {author} {\bibfnamefont {O.}~\bibnamefont {Seydi}}, \ and\ \bibinfo {author}
  {\bibfnamefont {G.}~\bibnamefont {Webb}},\ }\href@noop {} {\bibfield
  {journal} {\bibinfo  {journal} {Infect. Dis. Model.}\ } (\bibinfo {year}
  {2020})}\BibitemShut {NoStop}%
\bibitem [{\citenamefont {Arino}\ and\ \citenamefont
  {Portet}(2020)}]{arino2020simple}%
  \BibitemOpen
  \bibfield  {author} {\bibinfo {author} {\bibfnamefont {J.}~\bibnamefont
  {Arino}}\ and\ \bibinfo {author} {\bibfnamefont {S.}~\bibnamefont {Portet}},\
  }\href@noop {} {\bibfield  {journal} {\bibinfo  {journal} {Infect. Dis.
  Model.}\ } (\bibinfo {year} {2020})}\BibitemShut {NoStop}%
\bibitem [{\citenamefont {Brauer}\ \emph {et~al.}(2019)\citenamefont {Brauer},
  \citenamefont {Castillo-Chavez},\ and\ \citenamefont
  {Feng}}]{brauer2019mathematical}%
  \BibitemOpen
  \bibfield  {author} {\bibinfo {author} {\bibfnamefont {F.}~\bibnamefont
  {Brauer}}, \bibinfo {author} {\bibfnamefont {C.}~\bibnamefont
  {Castillo-Chavez}}, \ and\ \bibinfo {author} {\bibfnamefont {Z.}~\bibnamefont
  {Feng}},\ }\href@noop {} {\emph {\bibinfo {title} {Mathematical Models in
  Epidemiology}}}\ (\bibinfo  {publisher} {Springer},\ \bibinfo {year}
  {2019})\BibitemShut {NoStop}%
\bibitem [{\citenamefont {Van~Mieghem}(2014)}]{van2014exact}%
  \BibitemOpen
  \bibfield  {author} {\bibinfo {author} {\bibfnamefont {P.}~\bibnamefont
  {Van~Mieghem}},\ }\href@noop {} {\bibfield  {journal} {\bibinfo  {journal}
  {arXiv preprint arXiv:1402.1731}\ } (\bibinfo {year} {2014})}\BibitemShut
  {NoStop}%
\bibitem [{\citenamefont {G{\'o}mez}\ \emph {et~al.}(2010)\citenamefont
  {G{\'o}mez}, \citenamefont {Arenas}, \citenamefont {Borge-Holthoefer},
  \citenamefont {Meloni},\ and\ \citenamefont {Moreno}}]{gomez2010discrete}%
  \BibitemOpen
  \bibfield  {author} {\bibinfo {author} {\bibfnamefont {S.}~\bibnamefont
  {G{\'o}mez}}, \bibinfo {author} {\bibfnamefont {A.}~\bibnamefont {Arenas}},
  \bibinfo {author} {\bibfnamefont {J.}~\bibnamefont {Borge-Holthoefer}},
  \bibinfo {author} {\bibfnamefont {S.}~\bibnamefont {Meloni}}, \ and\ \bibinfo
  {author} {\bibfnamefont {Y.}~\bibnamefont {Moreno}},\ }\href@noop {}
  {\bibfield  {journal} {\bibinfo  {journal} {Europhys. Lett.}\ }\textbf
  {\bibinfo {volume} {89}},\ \bibinfo {pages} {38009} (\bibinfo {year}
  {2010})}\BibitemShut {NoStop}%
\bibitem [{\citenamefont {Bonacich}(1972)}]{bonacich1972factoring}%
  \BibitemOpen
  \bibfield  {author} {\bibinfo {author} {\bibfnamefont {P.}~\bibnamefont
  {Bonacich}},\ }\href@noop {} {\bibfield  {journal} {\bibinfo  {journal} {J.
  Math. Sociol.}\ }\textbf {\bibinfo {volume} {2}},\ \bibinfo {pages} {113}
  (\bibinfo {year} {1972})}\BibitemShut {NoStop}%
\bibitem [{\citenamefont {Gatto}\ \emph {et~al.}(2012)\citenamefont {Gatto},
  \citenamefont {Mari}, \citenamefont {Bertuzzo}, \citenamefont {Casagrandi},
  \citenamefont {Righetto}, \citenamefont {Rodriguez-Iturbe},\ and\
  \citenamefont {Rinaldo}}]{gatto2012generalized}%
  \BibitemOpen
  \bibfield  {author} {\bibinfo {author} {\bibfnamefont {M.}~\bibnamefont
  {Gatto}}, \bibinfo {author} {\bibfnamefont {L.}~\bibnamefont {Mari}},
  \bibinfo {author} {\bibfnamefont {E.}~\bibnamefont {Bertuzzo}}, \bibinfo
  {author} {\bibfnamefont {R.}~\bibnamefont {Casagrandi}}, \bibinfo {author}
  {\bibfnamefont {L.}~\bibnamefont {Righetto}}, \bibinfo {author}
  {\bibfnamefont {I.}~\bibnamefont {Rodriguez-Iturbe}}, \ and\ \bibinfo
  {author} {\bibfnamefont {A.}~\bibnamefont {Rinaldo}},\ }\href@noop {}
  {\bibfield  {journal} {\bibinfo  {journal} {PNAS}\ }\textbf {\bibinfo
  {volume} {109}},\ \bibinfo {pages} {19703} (\bibinfo {year}
  {2012})}\BibitemShut {NoStop}%
\bibitem [{\citenamefont {Boccaletti}\ \emph {et~al.}(2006)\citenamefont
  {Boccaletti}, \citenamefont {Latora}, \citenamefont {Moreno}, \citenamefont
  {Chavez},\ and\ \citenamefont {Hwang}}]{boccaletti2006complex}%
  \BibitemOpen
  \bibfield  {author} {\bibinfo {author} {\bibfnamefont {S.}~\bibnamefont
  {Boccaletti}}, \bibinfo {author} {\bibfnamefont {V.}~\bibnamefont {Latora}},
  \bibinfo {author} {\bibfnamefont {Y.}~\bibnamefont {Moreno}}, \bibinfo
  {author} {\bibfnamefont {M.}~\bibnamefont {Chavez}}, \ and\ \bibinfo {author}
  {\bibfnamefont {D.-U.}\ \bibnamefont {Hwang}},\ }\href@noop {} {\bibfield
  {journal} {\bibinfo  {journal} {Phys. Rep.}\ }\textbf {\bibinfo {volume}
  {424}},\ \bibinfo {pages} {175} (\bibinfo {year} {2006})}\BibitemShut
  {NoStop}%
\bibitem [{\citenamefont {Erdos}\ and\ \citenamefont
  {R{\'e}nyi}(1960)}]{erdos1960evolution}%
  \BibitemOpen
  \bibfield  {author} {\bibinfo {author} {\bibfnamefont {P.}~\bibnamefont
  {Erdos}}\ and\ \bibinfo {author} {\bibfnamefont {A.}~\bibnamefont
  {R{\'e}nyi}},\ }\href@noop {} {\bibfield  {journal} {\bibinfo  {journal}
  {Publ. Math. Inst. Hung. Acad. Sci}\ }\textbf {\bibinfo {volume} {5}},\
  \bibinfo {pages} {17} (\bibinfo {year} {1960})}\BibitemShut {NoStop}%
\bibitem [{\citenamefont {Jury}(1964)}]{jury1964theory}%
  \BibitemOpen
  \bibfield  {author} {\bibinfo {author} {\bibfnamefont {E.~I.}\ \bibnamefont
  {Jury}},\ }\href@noop {} {\emph {\bibinfo {title} {Theory and Application of
  the z-Transform Method}}}\ (\bibinfo  {publisher} {Wiley},\ \bibinfo {year}
  {1964})\BibitemShut {NoStop}%
\bibitem [{\citenamefont {Ogata}(1995)}]{ogata1995discrete}%
  \BibitemOpen
  \bibfield  {author} {\bibinfo {author} {\bibfnamefont {K.}~\bibnamefont
  {Ogata}},\ }\href@noop {} {\emph {\bibinfo {title} {Discrete-time control
  systems}}},\ Vol.~\bibinfo {volume} {2}\ (\bibinfo  {publisher} {Prentice
  Hall Englewood Cliffs, NJ},\ \bibinfo {year} {1995})\BibitemShut {NoStop}%
\end{thebibliography}%


%

\appendix

\section{Stability condition for the disease-free equilibrium in the compartmental model}\label{A:Stab_compart}

As is given in the main text, the eigenvalues of the Jacobian at the disease-free state of the compartmental model are obtained from the determinant $\det (\mathbf{J}_{\text{DF}} - \lambda\mathbf{I})$, that is 
\begin{eqnarray}
&\begin{bmatrix}
  -\lambda &
    0 &
    -\alpha & -\beta & 0 \\[1ex] 
  0 &
 -\gamma - \lambda & \alpha & \beta & 0 \\[1ex]
  0 & \gamma & -\sigma-\mu-\lambda & 0
    & 0 \\[1ex]
    0 & 0 & \sigma & -\mu - \lambda & 0\\[1ex]
    0 & 0 & \mu & \mu & -\lambda\\
\end{bmatrix} \nonumber\\
&= \lambda^2 \det \begin{bmatrix}
 -\gamma -\lambda & \alpha & \beta \\[1ex]
  \gamma & -\sigma-\mu - \lambda & 0\\[1ex]
 0 & \sigma & -\mu -\lambda
    \end{bmatrix}.
\label{eq:SEAIR_compart_Jacob}
\end{eqnarray}
Besides the two trivial eigenvalues $\lambda=0$, the remaining three are the roots of the polynomial which is obtained by expanding the last determinant
\begin{eqnarray}
\mathcal{R}(\lambda)
    &=& (-\mu-\lambda)\left[(-\gamma - \lambda)(-\sigma-\mu-\lambda) -\alpha\gamma\right] \nonumber\\ 
    &+& \beta\gamma\sigma.\label{eq:char_func_compart}
\end{eqnarray}
The cubic polynomial in $\lambda$ in the last equation can be written in the form
\begin{equation}
    -\mathcal{R}(\lambda) = \lambda^3 + a_2\lambda^2 + a_1\lambda + a_0, \label{eq:poly_3}
\end{equation}
where the coefficients are
\begin{eqnarray}
    a_2 &=& \gamma + \sigma + 2\mu, \nonumber\\
    a_1 &=& \mu\sigma + \gamma\sigma + \mu^2 + 2\gamma\mu - \alpha\gamma, \nonumber \\
    a_0 &=& \gamma(\mu^2 + \mu\sigma - \alpha\mu - \beta\sigma).
    \label{eq:polynomial_coeff_compart}
\end{eqnarray}
By the Routh-Hurwitz criterion \cite{brauer2019mathematical}, the roots of the third order polynomial of the form (\ref{eq:poly_3}) will have negative real parts if and only if $a_2>0$ and $a_2 a_1>a_0>0$. Since $a_2>0$, the condition $a_0>0$ is equivalent to
\begin{equation}
    \mu(\mu + \sigma) > \alpha\mu + \beta\sigma. \label{eq:a_0>0}
\end{equation}
We should also verify that $a_2a_1>a_0$ is satisfied, which after multiplication of the respective values in (\ref{eq:polynomial_coeff_compart}) will result in
\begin{eqnarray}
    && \gamma\mu\sigma +\gamma^2\sigma+\gamma\mu^2 + 2\gamma^2\mu -\alpha\gamma^2 \nonumber\\ && + \mu\sigma^2 + \gamma\sigma^2 + \mu^2\sigma + 2\gamma\mu\sigma-\alpha\gamma\sigma \nonumber\\
    && +2\mu^2\sigma+2\gamma\mu\sigma+2\mu^3 + 4\gamma\mu^2 - 2\gamma\alpha\mu\nonumber\\
    && > \gamma\mu^2 +\gamma\mu\sigma -\gamma\alpha\mu - \gamma\beta\sigma.
\end{eqnarray}
By algebraic manipulation and rearrangement of the terms, the last inequality becomes
\begin{eqnarray}
    && 4\gamma\mu\sigma +\gamma^2\sigma+4\gamma\mu^2 + 2\gamma^2\mu \nonumber\\ && -\alpha\gamma^2  + \mu\sigma^2 + \gamma\sigma^2 + 3\mu^2\sigma  \nonumber \\
    && -\alpha\gamma\sigma +2\mu^3 -\gamma\alpha\mu + \gamma\beta\sigma > 0.\label{eq:a_2_a_1>a_0}
\end{eqnarray}
Now, from the condition (\ref{eq:a_0>0}) one has the inequality
\begin{equation}
    \mu + \sigma > \alpha,
\end{equation}
which implies the following relationships for the terms with minus sign before them in (\ref{eq:a_2_a_1>a_0})
\begin{eqnarray}
    \gamma^2(\mu + \sigma) &>& \gamma^2\alpha, \nonumber\\
    \gamma\sigma (\mu + \sigma) &>&\gamma\sigma\alpha, \nonumber\\
    \gamma\mu(\mu + \sigma) &>& \gamma\mu\alpha.
\end{eqnarray}
By using the last three inequalities in (\ref{eq:a_2_a_1>a_0}), one will obtain that only positive terms will remain at the left hand side, which means that it is satisfied. Thus by the Routh-Hurwitz criterion the nontrivial eigenvalues of the Jacobian have negative real parts if and only if (\ref{eq:a_0>0}) holds.

It is worth noting that disease-free state becomes unstable when the condition (\ref{eq:a_0>0}) is not satisfied. In that case $a_0 < 0$, which from the equation (\ref{eq:poly_3}) implies that
\begin{equation}
    \mathcal{R}(0) = -a_0 > 0.
\end{equation}
Since the sign of the highest term of the characteristic polynomial $\mathcal{R}(\lambda)$ is negative, it means that the cubic parabola is decreasing towards $-\infty$, when $\lambda \to \infty$. Thus, there must be a real root of the polynomial with positive sign, because the curve intersects the horizontal axis for some $\lambda > 0$. Thus, the root responsible for the instability of the disease-free equilibrium is real and positive one. This observation is important in the analysis of the equilibria in discrete-time models.

\section{Endemic equilibrium for random regular graph}\label{A:Rand_reg_endemic}

For determination of the population remaining unaffected by the epidemic in disease spreading on random regular graph, we follow the same approach as in the compartmental model. To proceed in that spirit, first sum up the first four equations in the system (\ref{eq:SEAIR_random_reg}) and obtain
\begin{eqnarray}
    && p_{S}(n+1) + p_{E}(n+1) + p_{A}(n+1) + p_{I}(n+1) \nonumber \\ &=& p_{S}(n) + p_{E}(n) + (1-\mu)\left[p_{A}(n) + p_{I}(n)\right]. \label{eq:SEAI_n_n+1}
\end{eqnarray}
We can sum the last relationship over all moments $n$, from the onset to the finish of the epidemic, and assume negligibly small initial probabilities of infected individuals. Then, due to cancellation of the respective terms it will be obtained
\begin{equation}
    p_S(0) - p_S(\infty) = \mu \sum_{n=0}^{\infty} \left[p_{A}(n) + p_{I}(n)\right], \label{eq:SEAIR_rand_reg_S_A_and_I}
\end{equation}
which corresponds to the equation (\ref{eq:delta_S_through_A_and_I}) of the compartmental model. 

Next, the fourth equation in (\ref{eq:SEAIR_random_reg}) is rewritten as \begin{equation}
    p_I(n+1) - p_I(n) = \sigma p_A(n) -\mu p_{I}(n). \label{eq:SEAIR_rand_reg_A_and_I}
\end{equation}
Summation of the infinite ladder of equations (\ref{eq:SEAIR_rand_reg_A_and_I}) and using $p_I(0) \approx 0 = p_I(\infty)$ leads to similar relationship between the probabilities of asymptomatic and infectious states as in the case of the compartmental model (\ref{eq:total_A_VS_I}),
\begin{equation}
    \sigma \sum_{n=0}^{\infty} p_{A}(n) = \mu \sum_{n=0}^{\infty} p_I(n). \label{eq:SEAIR_rand_reg_sum_A_and_I}
\end{equation}

The first equation in (\ref{eq:SEAIR_random_reg}) can be written as
\begin{equation}
     \left[\frac{p_S(n+1)}{p_S(n)}\right]^{1/k} = 1-\beta p_I(n) - \alpha p_A(n).
\end{equation}
If we take logarithm of the last equation, and use approximation $\alpha p_A(n) \ll 1$ and $\beta p_I(n) \ll 1$, that holds for weak spreading $\alpha \ll 1$ and $\beta \ll 1$, it will be obtained
\begin{equation}
     \frac{1}{k}\left[\ln p_S(n+1) - \ln p_S(n)\right] = -\beta p_I(n) - \alpha p_A(n).
\end{equation}
Summing the last relationship for all moments, after cancellations, results in
\begin{equation}
    \ln p_S(0) - \ln p_S(\infty) = k\alpha \sum_{n=0}^{\infty} p_{A}(n) + k\beta \sum_{n=0}^{\infty} p_{I}(n). \label{eq:SEAIR_rand_reg_log_S_A_and_I}
\end{equation}

Using the relationships (\ref{eq:SEAIR_rand_reg_S_A_and_I}), (\ref{eq:SEAIR_rand_reg_sum_A_and_I}) and (\ref{eq:SEAIR_rand_reg_log_S_A_and_I}) one can obtain an estimate of the number of unaffected individuals in epidemic spreading on random regular graphs from
\begin{equation}
  p_{S}(0) - p_{S}(\infty) = \frac{\mu(\mu + \sigma)}{k(\beta\sigma + \alpha\mu)} \ln\frac{p_{S}(0)}{p_{S}(\infty)}. \label{eq:final_state_suscept}
\end{equation}

\section{Stability of disease-free equilibrium for random regular graph} \label{A:Stab_rand_reg}

The characteristic polynomial of the Jacobian of the disease-free state in the discrete-time SEAIR model on random regular graph can be compactly written as
\begin{equation}
    -\mathcal{S}(\lambda) = \lambda^3 + b_2 \lambda^2 + b_1 \lambda + b_0. \label{eq:charact_poly_rand_reg}
\end{equation}
Its coefficients are related to those of the compartmental model 
(\ref{eq:polynomial_coeff_compart}) with
\begin{eqnarray} 
    b_0 &=& a_0 - a_1 + a_2 - 1, \nonumber\\
    b_1 &=& a_1 - 2a_2 + 3, \nonumber\\
    b_2 &=& a_2 - 3, \label{eq:polynomial_coeff_reg} 
\end{eqnarray}
where we remind that in the expressions (\ref{eq:polynomial_coeff_compart}) for the coefficients $a_0$, $a_1$ and $a_2$ one should use $k\alpha$ and $k\beta$ instead of $\alpha$ and $\beta$, respectively.

An equilibrium of a discrete-time dynamical system is linearly stable, if the modulus of the dominant eigenvalue of the associated Jacobian matrix does not exceed one. It means that all roots of the characteristic polynomial of the Jacobian are within the unit circle. For polynomials with real coefficients this is verified with the Jury test \cite{jury1964theory, ogata1995discrete}, which is discrete-time analogue to the Routh-Hurwitz criterion. According to the Jury test, the roots of the polynomial lie within the unit circle, if an only if the following four conditions are met
\begin{eqnarray}
    \mathcal{S}(1) &>& 0, \nonumber \\
    \mathcal{S}(-1) &<& 0, \nonumber \\
    |b_0| &<& 1, \nonumber \\
    |1 - b_0^2| &>& |b_0 b_2 - b_1|.
\end{eqnarray}
In the application of the test, the conditions above are checked in the given order and if one is not satisfied, than at least one root is outside the unit circle and the equilibrium is unstable. The verification of the first condition in the Jury test $\mathcal{S}(1) > 0$, by using the relationships (\ref{eq:polynomial_coeff_compart}) and (\ref{eq:polynomial_coeff_reg}) leads to the demand that $a_0 > 0$, which also appeared in the analysis of the compartmental case. This will result in the following inequality
\begin{equation}
    \mu(\mu + \sigma) > k (\alpha \mu + \beta \sigma),
\end{equation}
which if holds, also implies that the endemic equilibrium does not exist. The last inequality is similar to the respective one for the compartmental case, with only difference being the presence of the node degree $k$. It can be shown that the second and the third condition of the Jury test are satisfied once the first one holds. The verification of the fourth condition is very complex, since it involves several dozens of products of the parameters up to degree six. Thus, a numerical verification was applied by taking all combinations of 100 different equally spaced values for the parameters $\alpha, \beta, \gamma$ and $\sigma$ in the range $(0,1)$, while for $\mu$ in the range $1 - \sigma$ for each $\sigma$, because $1-\mu - \sigma$ is the probability for an asymptomatic person to remain so in the next time step. This procedure has shown that the fourth condition is satisfied if the first one is fulfilled as well. However, because theoretical verification is not complete without the fourth condition in the Jury test is analytically checked, the presentation of the proofs that the second and the third are satisfied is omitted for brevity. 

If one compares the characteristic polynomials for the compartmental (\ref{eq:char_func_compart_main}) and discrete-time model on random regular graph (\ref{eq:SEAIR_determinant_main}), by substituting $1 - \lambda$ in the latter with $-\lambda$ and taking $k=1$ will obtain the former. This implies that the roots of the compartmental model $\lambda_c$ are related with those of the discrete-time case $\lambda_d$ with
\begin{equation}
    \lambda_d = \lambda_c + 1.
    \label{eq:_poly_roots_relation}
\end{equation}
The last equation implies that the disease-free state of the discrete-time model becomes unstable due to existence of real eigenvalue of the Jacobian that is greater than one.

\section{Endemic equilibrium for complex network} \label{A:Complex_endemic}

For small contagiousness parameters $\alpha \ll 1$ and $\beta \ll 1$, one can approximate the probability that a susceptible individual will not receive the virus as
\begin{align}
    \prod_{j\in \mathcal{N}_i} \left[1-\alpha p_{A,j}(n)-\beta p_{I,j}(n)\right] \\ \nonumber \approx 1 - \alpha \sum_{j\in \mathcal{N}_i}p_{A,j}(n) - \beta \sum_{j\in \mathcal{N}_i}p_{I,j}(n). \label{eq:Prob_not_infected_approx}
\end{align}
Then the evolution of all probabilities can be compactly written as
\begin{eqnarray}
    \mathbf{p}_S(n + 1) &=& \mathbf{p}_S(n)\left[\mathbf{I} - \alpha \mathbf{A} \mathbf{p}_A(n)- \beta \mathbf{A} \mathbf{p}_I(n) \right],\nonumber \\
    \mathbf{p}_E(n + 1) &=& \mathbf{p}_S(n)\left[\alpha \mathbf{A} \mathbf{p}_A(n)+ \beta \mathbf{A} \mathbf{p}_I(n) \right] \nonumber \\ &+& (1-\gamma)\mathbf{I}\mathbf{p}_E(n),\nonumber \\
    \mathbf{p}_A(n + 1) &=& \gamma \mathbf{I}\mathbf{p}_E(n) + (1-\sigma - \mu)\mathbf{I}\mathbf{p}_A(n),\nonumber \\
    \mathbf{p}_I(n + 1) &=& \sigma \mathbf{I}\mathbf{p}_A(n) + (1-\mu)\mathbf{I}\mathbf{p}_I(n),\nonumber \\
    \mathbf{p}_R(n + 1) &=&  \mathbf{I}\mathbf{p}_R(n) + \mu\mathbf{I}\left[\mathbf{p}_A(n)+\mathbf{p}_I(n)\right]. \label{eq:SEAIR_comp_net_vectors}
\end{eqnarray}
We will follow the same technique as for the previous two scenarios. Summing up the first four equations in the last system will result in
\begin{eqnarray}
    && \mathbf{p}_{S}(n+1) + \mathbf{p}_{E}(n+1) + \mathbf{p}_{A}(n+1) + \mathbf{p}_{I}(n+1) \nonumber \\ &=& \mathbf{p}_{S}(n) + \mathbf{p}_{E}(n) + (1-\mu)\left[\mathbf{p}_{A}(n) + \mathbf{p}_{I}(n)\right].
\end{eqnarray}
Now, lets sum over all moments and use the fact that the probabilities of infected states at the beginning and ending of epidemic are vanishing. Then from the last relationship will be obtained
\begin{equation}
    \mathbf{p}_S(0) - \mathbf{p}_S(\infty) = \mu \sum_{n=0}^{\infty} \left[\mathbf{p}_{A}(n) + \mathbf{p}_{I}(n)\right]. \label{eq:SEAIR_complex_S_A_and_I}
\end{equation}
Rearrangement of the fourth equation in (\ref{eq:SEAIR_comp_net_vectors}) and summing over all moments will lead to result that generalizes (\ref{eq:SEAIR_rand_reg_sum_A_and_I})
\begin{equation}
    \sigma \sum_{n=0}^{\infty} \mathbf{p}_{A}(n) = \mu \sum_{n=0}^{\infty} \mathbf{p}_I(n). \label{eq:SEAIR_complex_sum_A_and_I}
\end{equation}
One can write the evolution equation of probability of the susceptible state for each node $i$ as
\begin{equation}
    \frac{p_{S,i}(n+1)}{p_{S,i}(n)} = \prod_{j \in \mathcal{N}_i} \left[1-\alpha p_{A,j}(n) - \beta p_{A,j}(n) \right].
\end{equation}
Further, take logarithm on both sides of the last equation and keep only leading terms in $\alpha$ and $\beta$ in the expansion of the logarithm of the multipliers  to obtain
\begin{equation}
    \ln\frac{p_{S,i}(n+1)}{p_{S,i}(n)} = - \alpha \sum_{j \in \mathcal{N}_i} p_{A,j}(n) - \beta\sum_{j \in \mathcal{N}_i} p_{I,j}(n). \label{eq:logPS_complex}
\end{equation}
Summing (\ref{eq:logPS_complex}) over all moments will result in
\begin{equation}
    \ln p_{S,i}(0) -\ln p_{S,i}(\infty) = \alpha \sum_{j \in \mathcal{N}_i} \sum_{n=0}^{\infty} p_{A,j}(n) + \beta \sum_{j \in \mathcal{N}_i} \sum_{n=0}^{\infty}p_{I,j}(n). \label{eq:sum_of_logs_complex}
\end{equation}
Denote with $\ln \mathbf{p}_S(n)$ the vector which components are the logarithms of probabilities of susceptible states $\ln p_{S,i}(n)$. Then, the relationship (\ref{eq:sum_of_logs_complex}) for all nodes
can be compactly written as
\begin{equation}
    \ln \mathbf{p}_S(0) - \ln \mathbf{p}_S(\infty) = \alpha \mathbf{A} \sum_{n=0}^{\infty} \mathbf{p}_{A}(n) + \beta \mathbf{A} \sum_{n=0}^{\infty} \mathbf{p}_{I}(n). \label{eq:logPs_sum}
\end{equation}
From one side, using (\ref{eq:SEAIR_complex_sum_A_and_I}) in (\ref{eq:SEAIR_complex_S_A_and_I}) will result in
\begin{equation}
    \mathbf{p}_{S}(0) - \mathbf{p}_{S}(\infty) = \mu (1 + \frac{\sigma}{\mu}) \sum_{n=0}^{\infty}  \mathbf{p}_{A}(n). \label{eq:pS_sum_with_pA}
\end{equation}
From another side, applying (\ref{eq:SEAIR_complex_sum_A_and_I}) in (\ref{eq:logPs_sum}) will lead to
\begin{equation}
    \ln \mathbf{p}_S(0) - \ln \mathbf{p}_S(\infty) = \left(\alpha + \beta\frac{\sigma}{\mu}\right) \mathbf{A} \sum_{n=0}^{\infty}  \mathbf{p}_{A}(n).\label{eq:ln_pS_sum_with_pA}
\end{equation}
The last two relationships are system of equations for determination of the vector of the probabilities of the susceptible state at the end of the epidemic and the infinite sum of the vectors of the asymptomatic states during the whole epidemic.

\section{Characteristic polynomial for the eigenvalues of the Jacobian of the discrete-time model}
\label{A:SEAIR_Charact_polynomial}

To obtain more compact expression for determination of the eigenvalues of the Jacobian (\ref{eq:SEAIR_Jacobian_complex_dis-free}), we will extensively use the Schur's determinant identity
\begin{equation}
    \det
\begin{bmatrix}
  \mathbf{Q} &
    \mathbf{R}\\[1ex] 
  \mathbf{S} &
 \mathbf{T} \end{bmatrix} = \det(\mathbf{T})\cdot \det(\mathbf{Q} - \mathbf{R}\mathbf{T}^{-1}\mathbf{S}). \label{eq:Schur's_det_identity}
\end{equation}
One should note that the identity does not need the matrices to be square and if at least one of the matrices $\mathbf{R}$ or $\mathbf{S}$ is zero, then one has simpler relationship
\begin{equation}
    \det
\begin{bmatrix}
  \mathbf{Q} &
    \mathbf{R}\\[1ex] 
  \mathbf{S} &
 \mathbf{T} \end{bmatrix} = \det(\mathbf{Q})\cdot \det(\mathbf{T}). \label{eq:Schur's_det_ident_simple}
\end{equation}
First we can assign the role of the bottom-right submatrix $\mathbf{T}$ in (\ref{eq:Schur's_det_identity}) to the bottom-right identity matrix in (\ref{eq:SEAIR_Jacobian_complex_dis-free}). Then one can note that to the respective submatrix $\mathbf{R}$ corresponds zero matrix and use (\ref{eq:Schur's_det_ident_simple}) instead to obtain 
\begin{widetext}
\begin{equation}
    \mathcal{T}(\lambda) = \det \left[(1-\lambda) \mathbf{I}\right]\cdot \det \begin{bmatrix}
     (1-\lambda)\mathbf{I} &
    \mathbf{0} &
    -\alpha \mathbf{A} & -\beta \mathbf{A}\\[1ex] 
  \mathbf{0} &
 (1-\gamma-\lambda)\mathbf{I} & \alpha \mathbf{A} \ & \beta \mathbf{A} \\[1ex]
  \mathbf{0} & \gamma \mathbf{I} & (1 - \sigma -\mu-\lambda) \mathbf{I} & \mathbf{0}
\\[1ex]
    \mathbf{0} & \mathbf{0} & \sigma\mathbf{I} & (1 - \mu-\lambda) \mathbf{I} \end{bmatrix}
\end{equation}
\end{widetext}
By repeating the same procedure one more time with taking top-left submatrix $(1-\lambda)\mathbf{I}$ as the submatrix $\mathbf{Q}$ in the Schur's determinant identity, and observing that now the submatrix $\mathbf{S}$ is zero, one can obtain that
\begin{eqnarray}
        &&\mathcal{T}(\lambda) = \left\{\det \left[(1-\lambda) \mathbf{I}\right]\right\}^2\nonumber \nonumber\\ &&\cdot \det \begin{bmatrix}
     (1-\gamma-\lambda)\mathbf{I} & \alpha \mathbf{A} & \beta \mathbf{A} \\[1ex]
 \gamma \mathbf{I} & (1 - \sigma -\mu-\lambda) \mathbf{I} & \mathbf{0}
\\[1ex]
 \mathbf{0} & \sigma \mathbf{I}& (1 - \mu-\lambda) \mathbf{I} \end{bmatrix}.
\end{eqnarray}
To simplify notation one could first stop repetitive writing of the part which contains the trivial eigenvalue $\lambda = 1$ which has multiplicity $2N$, and focus on the remaining. Take the submatrix $\mathbf{T}=(1-\mu - \lambda)\mathbf{I}$ which determinant contains trivial eigenvalues $\lambda = 1 - \mu$ and respectively the remaining submatrices $\mathbf{Q}$, $\mathbf{R}$ and $\mathbf{S}$. We note that $1 - \mu$ are not eigenvalues of the Jacobian, since in expanding the determinants as polynomial, the terms corresponding to $1-\mu-\lambda$ that appear in $\mathbf{T}$ will cancel with the same terms which will appear in the denominator in the remaining determinant as will be seen below. From the last determinant let us first consider the submatrix that corresponds to the product $\mathbf{R}\mathbf{T}^{-1}\mathbf{S}$ in the Schur's identity (\ref{eq:Schur's_det_identity}).  By using the properties of the inverse matrix one can obtain first 
\begin{equation}
     \begin{bmatrix}
     (1-\mu-\lambda)\mathbf{I}
     \end{bmatrix} ^ {-1}\cdot
     \begin{bmatrix}
     \mathbf{0} & \sigma\mathbf{I}
     \end{bmatrix} = \begin{bmatrix}
     \mathbf{0} & \frac{\sigma}{1-\mu-\lambda}\mathbf{I}
     \end{bmatrix}.
\end{equation}
Then it follows that
\begin{equation}
    \begin{bmatrix}
     \beta\mathbf{A} \\[1ex]
     \mathbf{0}
     \end{bmatrix}\cdot
     \begin{bmatrix}
     \mathbf{0} & \frac{\sigma}{1-\mu-\lambda}\mathbf{I}
     \end{bmatrix} = 
     \begin{bmatrix} \mathbf{0} & \frac{\beta\sigma}{1-\mu-\lambda}\mathbf{A}\\[1ex]
     \mathbf{0}  & \mathbf{0} 
    \end{bmatrix}.
\end{equation}
Now, the part of the characteristic polynomial which contains the nontrivial eigenvalues is
\begin{eqnarray}
\mathcal{U}(\lambda) &=& \det\left\{ \begin{bmatrix}
     (1-\gamma-\lambda)\mathbf{I} & \alpha \mathbf{A}  \\[1ex]
 \gamma \mathbf{I} & (1 - \sigma -\mu-\lambda) \mathbf{I} 
 \end{bmatrix} - \begin{bmatrix} \mathbf{0} & \frac{\beta\sigma}{1-\mu-\lambda}\mathbf{A}\\[1ex]
     \mathbf{0}  & \mathbf{0} 
    \end{bmatrix} \right\} \nonumber\\ &=& \det \begin{bmatrix}
     (1-\gamma-\lambda)\mathbf{I} & \left(\alpha - \frac{\beta\sigma}{1-\mu-\lambda} \right) \mathbf{A}  \\[1ex]
 \gamma \mathbf{I} & (1 - \sigma -\mu-\lambda) \mathbf{I} 
 \end{bmatrix}.
\end{eqnarray}

We can apply the Schur's identity again. First observe the matrix product that corresponds to the $\mathbf{R}\mathbf{T}^{-1}\mathbf{S}$ term in (\ref{eq:Schur's_det_identity})
\begin{eqnarray}
    && \left(\alpha - \frac{\beta\sigma}{1-\mu-\lambda} \right) \mathbf{A} \cdot \left[(1 - \mu -\sigma-\lambda) \mathbf{I}\right]^{-1} \cdot \gamma \mathbf{I} \nonumber\\ &=& \frac{\gamma}{1-\sigma-\mu-\lambda}\left(\alpha - \frac{\beta\sigma}{1-\mu-\lambda} \right) \mathbf{A}.
\end{eqnarray}
After simplification of the scalar at the right-hand side of the last relationship and subtract the respective matrices in the form $\mathbf{Q}-\mathbf{R}\mathbf{T}^{-1}\mathbf{S}$ from (\ref{eq:Schur's_det_identity}) one will obtain the following characteristic polynomial of the eigenvalues
\begin{widetext}
\begin{equation}
    \mathcal{U}(\lambda) = \det\left[(1-\sigma-\mu-\lambda)\textbf{I}  \right] \cdot \det\left[(1-\gamma-\lambda)\mathbf{I} - \frac{\gamma[\alpha(1-\mu-\lambda) - \beta\sigma]}{(1-\sigma-\mu-\lambda)(1-\mu-\lambda)}\mathbf{A} \right]. \label{eq:next_to_last_det}
\end{equation}
\end{widetext}
Again, the first determinant has trivial eigenvalues $\lambda = 1-\sigma-\mu$ with multiplicity $N$ as well and the nontrivial ones are contained in the second determinant. By observing the second determinant in (\ref{eq:next_to_last_det}) one can note that in the denominator multiplying the adjacency matrix appear terms $1-\mu-\lambda$ and $1-\sigma-\mu-\lambda$. Expansion of the determinants as polynomials will result in cancellation of those terms in the denominators with the respective ones in the determinants $\det\left[(1-\mu-\lambda)\textbf{I}  \right]$ and $\det\left[(1-\sigma-\mu-\lambda)\textbf{I}  \right]$. Finally, the characteristic polynomial resulting from the last nontrivial determinant will not change if one multiplies it with a constant. So, a more convenient form of the last determinant, and the respective characteristic polynomial is
\begin{equation}
    \mathcal{V}(\lambda) = \det\left[\frac{(1-\gamma-\lambda)(1-\sigma-\mu-\lambda)(1-\mu-\lambda)}{\gamma[\alpha(1-\mu-\lambda) - \beta\sigma]}\mathbf{I} - \mathbf{A} \right].
\end{equation}

\section{Stability of the endemic equilibrium in disease spreading on complex networks} \label{A:Endem_stability}

Since the Jacobian of the endemic and of the disease-free equilibrium differ only in the presence of the matrix $\mathbf{\Sigma}$, the characteristic equation will have the same form for both cases. However, it was previously obtained that the leading eigenvalue of the Jacobian of the disease-free equilibrium $\lambda_{\max}$ depends on the leading one of the adjacency matrix $\Lambda_{max}$. Accordingly, for the endemic state the dependence will be on the leading eigenvalue $L_{\max}$ of the matrix product $\mathbf{\Sigma A}$. We will verify that this eigenvalue is related with that of the adjacency matrix as $L_{\max} < p_{S,\max}\Lambda_{max}$, where $p_{S,\max} = \max p_{S,i}(\infty)$, is the maximum of the probabilities of susceptible states at the end of the epidemic. To prove that, denote with $\mathbf{x}$ the unit eigenvector of $\mathbf{\Sigma A}$, corresponding to $L_{\max}$, or $\mathbf{\Sigma A x} = L_{\max} \mathbf{x}$. Let $\Lambda_i$ and $\mathbf{u}_i$ are the eigenvalues and the respective orthogonal basis vectors corresponding to the adjacency matrix. The vector $\mathbf{x}$ in the basis $\mathbf{u}_i$ is given as
\begin{equation}
    \mathbf{x} = \sum_{i=1}^{N} a_i \mathbf{u}_i,
\end{equation}
where $\sum a_i^2 = 1$ because $\mathbf{x}$ is unit vector. Then, multiplying the matrix $\mathbf{A}$ with $\mathbf{x}$ will result in some vector
\begin{equation}
    \mathbf{y} = \mathbf{A x} = \sum_{i=1}^{N} a_i \Lambda_i \mathbf{u}_i,\label{eq:y_equals_A_x}
\end{equation}
Due to the orthonormality of the basis $\mathbf{u}_i^{\text{T}}\mathbf{u}_j = \delta_{i,j}$, the squared magnitude of $\mathbf{y}$ reads
\begin{equation}
    \mathbf{y}^\text{T}\mathbf{y} =\sum_{i=1}^{N} a_i^2 \Lambda_i^2,
\end{equation}
which can be bounded as
\begin{equation}
    \mathbf{y}^\text{T}\mathbf{y} \leq \Lambda_{\max}^2\sum_{i=1}^{N} a_i^2 = \Lambda_{\max}^2. \label{eq:y_magnitude}
\end{equation}
This means that the vector $\mathbf{y}$ has length not bigger than $\Lambda_{\max}$. In connected network each node will be infected with nonzero probability, and thus $p_{S, i}(\infty) < 1$. Then the matrix $\mathbf{\Sigma}$ is symmetric  positive semi-definite, and all its eigenvalues are strictly less than one. Let us now express the vector $\mathbf{y}$ in the orthonormal basis $\mathbf{v}_i$ of the matrix $\mathbf{\Sigma}$ 
\begin{equation}
    \mathbf{y} = \sum_{i=1}^N b_i \mathbf{v}_i.
\end{equation}
Then the vector $\mathbf{\Sigma y}$ can be expressed as
\begin{equation}
    \mathbf{\Sigma y} = \sum_{i=1}^N b_i p_{S,i}(\infty) \mathbf{v}_i,
\end{equation}
since $\mathbf{\Sigma}$ is diagonal matrix with eigenvalues $p_{S,i}(\infty)$. The squared magnitude of $\mathbf{\Sigma y}$ is bounded as
\begin{eqnarray}
    &\mathbf{y}^{\text{T}}\mathbf{\Sigma}^{\text{T}}\mathbf{\Sigma y}= \sum_{i=1}^N b_i^2 p_{S,i}(\infty)^2\nonumber \\ &< p_{S,\max}^2 \sum_{i=1}^N b_i^2 = p_{S,\max}^2 |\mathbf{y}|^2. \label{eq:Sigma_y_magnitude}
\end{eqnarray}
Now, combining (\ref{eq:y_equals_A_x}), (\ref{eq:y_magnitude}) and (\ref{eq:Sigma_y_magnitude}) will result in
\begin{equation}
    L_{\max}^2 \mathbf{x}^{\text{T}}\mathbf{x} =  \mathbf{x}^{\text{T}}\mathbf{A}^{\text{T}}\mathbf{\Sigma}^{\text{T}}\mathbf{\Sigma A x} < p_{S,\max}^2\Lambda_{\max}^2.
\end{equation}
Thus, we have just bounded the leading eigenvalue of the matrix $\mathbf{\Sigma A}$ as
\begin{equation}
L_{\max} < p_{S,\max}\Lambda_{\max}.\label{eq:L_and_Lambda}
\end{equation}

Recall that in the stability analysis of the endemic equilibrium one has the matrix product $\mathbf{\Sigma A}$ instead of $\mathbf{A}$ which is used for the disease-free state. So, the stability of the endemic equilibrium depends on $L_{\max}$ as the other case depends on $\Lambda_{\max}$. Correspondingly, the endemic equilibrium will be linearly stable, once the following inequality holds [refer to the respective condition (\ref{eq:stab_dis-free_complex})]
\begin{equation}
    \mu(\mu+\sigma) > L_{\max}(\alpha \mu +\beta\sigma). \label{eq:Endem_stab_complex}
\end{equation}
Now, consider the system of transcendental equations (\ref{eq:Susc_endemic_complex}) and use the fact that $\mathbf{p}_A$ is principal eigenvector of the adjacency matrix $\mathbf{A}$, or $\mathbf{A p}_A = \Lambda_{\max}\mathbf{p}_A$. By algebraic manipulations, from the system (\ref{eq:Susc_endemic_complex}) it can be shown that for each component of the susceptible probability vector holds relationship similar to (\ref{eq:final_fract_suscept})
\begin{equation}
  p_{S,i}(0) - p_{S,i}(\infty)  = \frac{\mu(\mu + \sigma)}{\Lambda_{\max}(\beta\sigma + \alpha\mu)} \left[\ln p_{S,i}(0) - \ln p_{S,i}(\infty)\right], \label{eq:final_p_s_i_complex}
\end{equation}
from which one has
\begin{equation}
    \frac{\mu(\mu + \sigma)}{\beta\sigma + \alpha\mu} = \Lambda_{\max}\frac{p_{S,i}(0) - p_{S,i}(\infty)}{\ln p_{S,i}(0) - \ln p_{S,i}(\infty)}. \label{eq:Endem_complex_ps_ln_ps_fraction}
\end{equation}
Combining the endemic equilibrium stability condition (\ref{eq:Endem_stab_complex}) with the last relationship (\ref{eq:Endem_complex_ps_ln_ps_fraction}) will result in
\begin{equation}
    \Lambda_{\max}\frac{p_{S,i}(0) - p_{S,i}(\infty)}{\ln p_{S,i}(0) - \ln p_{S,i}(\infty)} > L_{\max}.
\end{equation}
Rearranging the terms in the last inequality will result in more convenient form
\begin{equation}
    \Lambda_{\max}p_{S,i}(\infty)\frac{\frac{p_{S,i}(0)}{p_{S,i}(\infty)}-1}{\ln \frac{p_{S,i}(0)}{p_{S,i}(\infty)}} > L_{\max}. \label{eq:endem_Stab_complex_proved}
\end{equation}
The last inequality is satisfied since one can use (\ref{eq:L_and_Lambda}) and the fraction at the left-hand side is always greater than one. Thus, when an epidemic occurs such that small fraction of the population is affected, the respective endemic equilibrium is linearly stable.

\section{Eigenvalues and eigenvectors of the Jacobian at the disease-free state for epidemic spreading on complex networks}\label{A:eigenval_eignvec_Jacob}

Denote the eigenvectors of the Jacobian matrix in the disease-free equilibrium with $\mathbf{w} = [\mathbf{w}_S^{\text{T}}, \mathbf{w}_E^{\text{T}}, \mathbf{w}_A^{\text{T}}, \mathbf{w}_I^{\text{T}}, \mathbf{w}_R^{\text{T}}]^{\text{T}}$ where $\mathbf{w}_S$, $\mathbf{w}_E$, $\mathbf{w}_A$, $\mathbf{w}_I$ and $\mathbf{w}_R$ are the column vectors which correspond to probabilities of the states S, E, A, I and R respectively. Then, the eigenvalue equation for the Jacobian $\mathbf{J w} = \lambda \mathbf{w}$ in more detailed form is 
\begin{widetext}
\begin{equation}
    \begin{bmatrix}
  \mathbf{I} &
    \mathbf{0} &
    -\alpha \mathbf{A} & -\beta \mathbf{A} & \mathbf{0} \\[1ex] 
  \mathbf{0} &
 (1-\gamma)\mathbf{I} & \alpha \mathbf{A} \ & \beta \mathbf{A} & \mathbf{0} \\[1ex]
  \mathbf{0} & \gamma \mathbf{I} & (1 - \sigma -\mu) \mathbf{I} & \mathbf{0}
    & \mathbf{0} \\[1ex]
    \mathbf{0} & \mathbf{0} & \sigma\mathbf{I} & (1 - \mu) \mathbf{I} & \mathbf{0}\\[1ex]
    \mathbf{0} & \mathbf{0} & \mu \mathbf{I}& \mu \mathbf{I}& \mathbf{I}\\
\end{bmatrix} \cdot \begin{bmatrix}
  \mathbf{w}_S \\ \mathbf{w}_E \\ \mathbf{w}_A \\ \mathbf{w}_I \\ \mathbf{w}_R
\end{bmatrix} = \lambda \begin{bmatrix}
  \mathbf{w}_S \\ \mathbf{w}_E \\ \mathbf{w}_A \\ \mathbf{w}_I \\ \mathbf{w}_R
\end{bmatrix}. \label{eq:eigeneq_Jacob}
\end{equation}
\end{widetext}
From the fourth row in (\ref{eq:eigeneq_Jacob}), which corresponds to the infectious state, one can obtain that
\begin{equation}
    \sigma \mathbf{w}_A + (1- \mu) \mathbf{w}_I = \lambda \mathbf{w}_I,
\end{equation}
which can be rearranged into
\begin{equation}
    \sigma \mathbf{w}_A = (\mu + \lambda - 1) \mathbf{w}_I. \label{eq:A_and_I_eigenv}
\end{equation}
The last equation relates the magnitudes of the vectors $\mathbf{w}_A$ and $\mathbf{w}_I$ and shows that they are collinear. In similar manner, from the last row in (\ref{eq:eigeneq_Jacob}), one can show that the vector $\mathbf{w}_R$ is collinear with the previous ones and moreover
\begin{equation}
    \frac{\mu(\mu + \lambda + \sigma + 1)}{\mu + \lambda + 1} \mathbf{w}_A = (\lambda - 1) \mathbf{w}_R. \label{eq:A_and_R_eigenv}
\end{equation}
Likewise, from the third row in (\ref{eq:eigeneq_Jacob}) it follows that the exposed probability vector $\mathbf{w}_E$ is also collinear to the previous ones, or more precisely 
\begin{equation}
    \gamma \mathbf{w}_E = (\mu + \lambda + \sigma - 1) \mathbf{w}_A. \label{eq:A_and_E_eigenv}
\end{equation}
Now consider the second row in (\ref{eq:eigeneq_Jacob}), from which one has
\begin{equation}
    \alpha \mathbf{A w}_A + \beta \mathbf{A w}_I = (\lambda + \gamma - 1) \mathbf{w}_E, 
\end{equation}
which after using (\ref{eq:A_and_I_eigenv}) and (\ref{eq:A_and_R_eigenv}) will result in
\begin{equation}
    \left(\alpha + \frac{\beta\sigma}{\mu + \lambda - 1} \right) \mathbf{A w}_A = (\lambda + \gamma - 1)\frac{(\mu + \lambda + \sigma - 1)}{\gamma} \mathbf{w}_A.
\end{equation}
The last relationship could be rearranged further as 
\begin{equation}
 \mathbf{A w}_A = \frac{(\lambda + \gamma - 1)(\mu + \lambda + \sigma - 1)(\mu + \lambda - 1)}{\gamma[\alpha(\mu + \lambda - 1) + \beta\sigma]} \mathbf{w}_A.\label{eq:A_eigenvec_selfcons}
\end{equation}
We have obtained eigenvalue equation for the adjacency matrix. Thus, every vector $\mathbf{w}_A$ must be eigenvector of the adjacency matrix $\mathbf{A}$. Since the eigenvalues of the adjacency matrix $\Lambda$ are independent on any dynamical process evolving on the network, it means that the eigenvalues of the Jacobian $\lambda$ must satisfy the relationship
\begin{equation}
    \Lambda = \frac{(\lambda + \gamma - 1)(\mu + \lambda + \sigma - 1)(\mu + \lambda - 1)}{\gamma[\alpha(\mu + \lambda - 1) + \beta\sigma]}. \label{eq:Two_lambdas}
\end{equation}
The last result relates the eigenvalues of the Jacobian with those of the adjacency matrix. By expanding the terms, one can see that it is cubic polynomial in $\lambda$, and thus for each eigenvalue $\Lambda$ one has three possibly different eigenvalues $\lambda$. Thus, $N$ eigenvalues of the adjacency matrix would generate $3N$ eigenvalues of the Jacobian. We remind that as is given in the Appendix \ref{A:SEAIR_Charact_polynomial}, there is one trivial eigenvalue $\lambda = 1$ with algebraic multiplicity $2N$. The eigenvectors corresponding to this eigenvalue are those that span the subspace consisting of susceptible or recovered states only and zeros at the remaining states. It can be easily verified from (\ref{eq:eigeneq_Jacob}), that each vector of the form \begin{equation}
    \mathbf{w}_{\text{S,R}} = [\mathbf{w}_S^{*\text{T}}, \mathbf{0}^{\text{T}}, \mathbf{0}^{\text{T}}, \mathbf{0}^{\text{T}}, \mathbf{w}_R^{*\text{T}}]^{\text{T}} \label{eq:S_R_subspace}
\end{equation} is eigenvector of the Jacobian.

Finally, from the first row in (\ref{eq:eigeneq_Jacob}) it follows that
\begin{equation}
    \left(\alpha + \frac{\beta\sigma}{\mu + \lambda - 1} \right) \mathbf{A w}_A = (1 - \lambda) \mathbf{w}_S,
\end{equation}
in which one can use (\ref{eq:A_eigenvec_selfcons}) to obtain
\begin{equation}
    \mathbf{w}_S = \frac{(\mu + \lambda + \sigma - 1)(\lambda + \gamma - 1)}{\gamma(1-\lambda)}\mathbf{w}_A. \label{eq:A_and_S_eigenv}
\end{equation}
Thus, the vector $\mathbf{w}_S$ is collinear with the rest as well. This implies that besides the vectors (\ref{eq:S_R_subspace}), the remaining eigenvectors of the Jacobian $\mathbf{w}$ consist of scaled copies of the eigenvectors of the adjacency matrix. More precisely, by using the relationships (\ref{eq:A_and_I_eigenv}), (\ref{eq:A_and_R_eigenv}), (\ref{eq:A_and_E_eigenv}) and (\ref{eq:A_and_S_eigenv}), the eigenvector $\mathbf{w}$ is
\begin{equation}
    \mathbf{w} = \begin{bmatrix}
  \mathbf{w}_S \\ \mathbf{w}_E \\ \mathbf{w}_A \\ \mathbf{w}_I \\ \mathbf{w}_R
\end{bmatrix} = \begin{bmatrix}\begin{array}{r}
      \frac{(\mu + \lambda + \sigma - 1)(\lambda + \gamma - 1)}{\gamma(1-\lambda)} \mathbf{w}_A \\
      \frac{\mu + \lambda + \sigma - 1}{\gamma} \mathbf{w}_A \\
      \mathbf{w}_A \\
      \frac{\sigma}{\mu + \lambda - 1} \mathbf{w}_A \\
      \frac{\mu(\mu + \lambda + \sigma - 1) }{(\mu + \lambda - 1)(\lambda - 1)} \mathbf{w}_A     
\end{array}
    \end{bmatrix} \label{eq:append_eigenvec_Jacob}
\end{equation}
Since $\mathbf{w}_A$ is eigenvector of the adjacency matrix, there are $3N$ eigenvectors of the form given in (\ref{eq:append_eigenvec_Jacob}). Together with $2N$ vectors of the form (\ref{eq:S_R_subspace}), they constitute a set of $5N$ eigenvectors. When they are orthogonal, with normalization one can obtain an orthonormal basis.

\end{document}